\documentclass{aa}  
\usepackage[cmyk]{xcolor}
\usepackage{upgreek}
\usepackage{subcaption}
\usepackage{orcidlink}
\usepackage{graphicx}
\usepackage{placeins}
\usepackage{txfonts}
\usepackage[normalem]{ulem}
\usepackage[switch]{lineno}

\begin{document}

   \title{Improved proper motion and gravity tests with PSR J1913+1102}

   \subtitle{}



      \author{Xueli Miao\inst{1,2}\orcidlink{https://orcid.org/0000-0003-1185-8937} 
      \and Paulo C. C. Freire\inst{2}\orcidlink{0000-0003-1307-9435}  
      \and Norbert Wex\inst{2}\orcidlink{https://orcid.org/0000-0003-4058-2837}
      \and Lingqi Meng\inst{2}\orcidlink{0000-0002-2885-568X} 
      \and Thomas M. Tauris\inst{3,2}\orcidlink{https://orcid.org/0000-0002-3865-7265}
      \and Junjie Zhao\inst{4}\orcidlink{https://orcid.org/0000-0002-9233-3683},
      \\
       Weiwei Zhu\inst{1,5}\orcidlink{https://orcid.org/0000-0001-5105-4058} 
      \and Robert Ferdman\inst{6}\orcidlink{https://orcid.org/0000-0002-2223-1235}
      \and Michael Kramer\inst{2,7}\orcidlink{https://orcid.org/0000-0002-4175-2271}
      \and Huanchen Hu\inst{8,2}\orcidlink{https://orcid.org/0000-0001-5105-4058}
      \and Lijing~Shao\inst{9,1,2}\orcidlink{https://orcid.org/0000-0002-1334-8853}
      \and Yanjun Guo\inst{1}, \\
      David~J.~Champion\inst{2}\orcidlink{https://orcid.org/0000-0003-1361-7723}
      \and Youling Yue\inst{1}\orcidlink{https://orcid.org/0000-0003-4415-2148}   
          }

   \institute{National Astronomical Observatories, Chinese Academy of Sciences, Beijing 100101, China\\
              \email{xlmiao@bao.ac.cn, zhuww@nao.cas.cn}
        \and Max-Planck-Institut f\"ur Radioastronomie, Auf dem H\"ugel 69, 53121 Bonn, Germany 
        \and Department of Materials and Production, Aalborg University, Fibigerstr{\ae}de 16, 9220 Aalborg, Denmark
        \and Institute for Gravitational Wave Astronomy, Henan Academy of Sciences, No. 228 Chongshili Road, Zhengzhou, 450046, Henan, China 
        \and Institute for Frontiers in Astronomy and Astrophysics, Beijing Normal University, Beijing 102206, China
        \and Faculty of Science, University of East Anglia, Norwich Research Park, Norwich NR4 7TJ, UK
        \and Jodrell Bank Centre for Astrophysics, Department of Physics and Astronomy, University of Manchester, M13 9PL, Manchester,UK
        \and Lohrmann Observatory, Technische Universit\"at Dresden, Mommsenstraße 13, 01062 Dresden, Germany
        \and Kavli Institute for Astronomy and Astrophysics, Peking University, Beijing 100871, China
 }

   \date{Received 24 February 2026 / Accepted 16 June 2026}

 
  \abstract
   {PSR J1913+1102 is a highly asymmetric double neutron star (DNS) system, making it an excellent laboratory for testing scalar-tensor gravity theories and a potential progenitor analogue of GW170817 that will merge in 470 Myr.}
   {To maximize the utility of this system for testing gravity theories, and to better understand its origin, we have carried out timing measurements with the aim of refining the proper motion and post-Keplerian (PK) parameters.}
   {We perform an updated timing analysis of the system by combining 13 years of data, consisting of historical Arecibo observations and the latest FAST measurements, and applying two different approaches to model the dispersion measure (DM) variations.}
   {The new timing analysis enables precise measurements of four PK parameters, leading to improved precision in mass estimates for the system. Assuming general relativity (GR) and modeling the DM variation with a Gaussian process, we obtain a three-fold improvement in the total mass $m_{\rm tot}=2.88948(20)\,{\rm M_{\odot}}$, and a nearly four-fold improvement in the pulsar mass $m_{\rm p}=1.599(8)\,{\rm M_{\odot}}$, the companion mass $m_{\rm c}=1.290(8)\,{\rm M_{\odot}}$ and  thus the mass ratio, $q=0.807(8)$.
   We also achieve an improved measurement of the system's proper motion, $\mu=7.71(25)\,{\rm mas\,yr^{-1}}$, enabling a more accurate determination of the non-intrinsic variation of the orbital period. This and the improved measurement of the variation of the orbital period lead to a refined measurement of the intrinsic variation of the orbital period, $\dot{P}_{\rm b}^{\rm intr}=-4.60(6)\times10^{-13}\,{\rm s\,s^{-1}}$, which is five times more precise than the previously published value and is fully consistent with the GR prediction for gravitational-wave damping. The combination of the improved mass measurements and the precise value of $\dot{P}_{\rm b}^{\rm intr}$ allows us to place a stringent constraint on the emission of dipolar gravitational waves and on the spontaneous scalarisation window around $1.6\,{\rm M_{\odot}}$. Our new measurement of the proper motion indicates that there is still room for improvement in the precision of this radiative test in the near future. The refined proper motion and mass measurements also provide tighter constraints on the final helium-star mass (immediately prior to its core collapse and formation of the second NS in a supernova), as well as on the magnitude and direction of the associated natal kick of the DNS system.}
   {}


   \maketitle
%

\section{Introduction}
Up to now, general relativity (GR) has been the most rigorously tested theory of gravity \citep[e.g.][and references therein]{Berti:2015,Will:2018book,Bambi:2024book}. However, under physically plausible conditions, GR predicts the existence of singularities, where the theory breaks down. Furthermore, it is a classical theory that is incompatible with quantum mechanics. These characteristics indicate that GR is not a complete theory of gravity. Moreover, alternative gravity theories exist that attempt to explain the ``missing mass'' problem in galaxies and the accelerating expansion of the Universe without introducing dark matter or dark energy \citep{Jain:2010AnPhy,Clifton:2012PhR}. Hence, rigorous tests of GR and alternative theories remain essential.

\citet{Wex:2014arXiv} introduced four gravity regimes to classify gravity tests, depending on the typical velocity of the masses, the strength of the gravitational field, and the radiative properties of gravity, and they are the quasi-stationary weak-field regime, the quasi-stationary strong-field regime, the highly-dynamical strong-field regime, and the radiation regime. Solar System experiments \citep[e.g.][and references therein]{Will:2018book,Ciufolini2024} have provided stringent tests of the quasi-stationary weak-field regime. 
However, some gravity theories reduce to GR in the quasi-stationary weak-field regime, but exhibit significant deviations from GR in the strong-field regime, as is the case for scalar-tensor gravity theories. Therefore, testing gravity across different regimes is essential for a comprehensive understanding of its fundamental nature.
The gravitational wave (GW) detections from LIGO/Virgo/KAGRA offer an opportunity to study gravity in the radiation regime and the highly dynamical strong-field regime \citep[e.g.][and references therein]{Sathyaprakash:2019,Colleoni:2024,Abbott:2025PhRvD}.
In contrast, the quasi-stationary strong-field regime is realized in well-separated binaries containing one or two compact objects.
In this context, precisely timed binary pulsar systems serve as natural laboratories for testing gravity in the quasi-stationary strong-field regime, for which they remain the only systems that can currently provide precision tests of gravity in this regime \citep[e.g.][and references therein]{Freire:2024,Hu:2024}.

\paragraph{Double neutron stars and tests of gravity theories:}The first discovered binary pulsar, PSR B1913+16---a double neutron star (DNS) system---provided the first evidence of GW radiation \citep{Hulse:1975ApJ,Taylor:1979,Weisberg:2016ApJ}.
Since then, binary pulsar systems have demonstrated their power as laboratories for testing gravity.
To date, more than 500 binary pulsar systems have been discovered\footnote{As of December 2025, 570 binary pulsar systems are listed in the ATNF Pulsar Catalogue.} \citep{Manchester:2005}. However, because DNS systems have undergone two supernova (SN) explosions, which greatly increase the probability of disruption, only 21 binary pulsars are confirmed PSR-NS systems\footnote{The ATNF Pulsar Catalogue includes 28 candidate PSR-NS systems, 21 of which have been confirmed. See also part 3 of \url{https://www3.mpifr-bonn.mpg.de/staff/pfreire/NS_masses.html}}.
In most PSR-NS systems, the detected pulsar is the first-born NS, which is mildly recycled, having been spun up through accretion from its companion before the second SN explosion.
Compared with normal pulsars, whose timing precision is limited by intrinsic timing noise and relatively larger time of arrival (ToA) measurement uncertainties, recycled pulsars are more stable rotators and provide more precise pulse ToAs. Their short spin periods allow averaging over many pulses within a given integration time, thereby reducing pulse jitter, i.e., stochastic variations in the shapes and arrival times of individual pulses, which contribute to ToA uncertainty.

DNS systems are ideal laboratories for probing gravity in the quasi-stationary strong-field regime. Their tight orbits and high compactness enable the measurement of several relativistic parameters in the motion of the pulsar, which are seen as perturbations of a pure Keplerian orbit. These effects are quantified by various post-Keplerian (PK) parameters, like the rate of periastron advance ($\dot{\omega}$), the variation of the orbital period ($\dot{P}_{\rm b}$), the Shapiro delay parameters ``range'' and ``shape'' ($r,\,s$) and the Einstein delay amplitude ($\gamma$) \citep{DD:1986,Damour:1992PrD}. 
In GR, to leading order, these PK parameters are functions of Keplerian parameters and the component masses of the binary. 
Consequently, if we assume the validity of GR, then the measurement of two PK parameters is enough to determine the masses of the pulsar and its companion.
The measurement of additional PK parameters provides redundancy, allowing stringent tests of the theory.
DNS systems thus represent some of the most precise astronomical laboratories for testing gravity. In particular, PSR J0737–3039A/B, in which both NSs were detected as radio pulsars, not only provides excellent tests of gravity but also enables the measurement of several relativistic effects for the first time \citep{Burgay:2003Natur,Kramer:2006Sci,Kramer:2021,Hu:2022A&A}.

\paragraph{PSR J1913+1102:}
This PSR-NS system was discovered by the Pulsar Arecibo L-band Feed Array survey (PALFA) in 2012 \citep{Lazarus:2016ApJ}. The detected pulsar has a spin period of $27.3\,$ms and resides in a mildly eccentric orbit ($e\sim0.09$) with an orbital period of 4.95 hours.
This system will merge in approximately $470$ Myr, making it a potential progenitor analogue of GW170817~\citep{Abbott:2017PhRvL,Abbott:2017ApJ,Abbott:2017ApJ_850, Cowperthwaite:2017ApJ}.
The relatively short spin period and low magnetic field suggest that it is the first-formed NS in this binary system, which was later spun up by accretion of matter from its companion \citep{Tauris:2023pbse.book}.
\citet{Ferdman:2020Natur} measured the three PK parameters: $\dot{\omega}$, $\gamma$ and $\dot{P}_{\rm b}$, based on observations covering MJD 56072-58747 (05.2012-09.2019). The precise measurement of $\dot{\omega}=5.6501\pm0.0007\,{\rm deg\,yr^{-1}}$ yielded a total mass of $m_{\rm tot} = 2.8887\pm 0.0006\,{\rm M_{\odot}}$ under the assumption of GR, making it the most massive DNS system known in our Galaxy\footnote{The DNS system PSR J0528+3529, recently discovered by the FAST Galactic Plane Pulsar Snapshot survey, has a total mass of $2.90(12)\,{\rm M_{\odot}}$ \citep{Wang:2025RAA}.
Some candidate DNS systems in globular clusters have even larger total masses: PSR~J1748$-$2446ao in Terzan 5 has a total mass of $3.166 \pm 0.024 \, \rm M_{\odot}$ \citep{Padmanabh:2024} and PSR~J0514$-$4002E in the globular cluster NGC 1851 has a total mass of $3.887 \pm 0.004 \, \rm M_{\odot}$ \citep{Barr:2024}. However, these systems were very likely formed dynamically, so not much can be inferred about the nature of their companions based on considerations from stellar evolution. In particular, PSR~J1748$-$2446ao system could also be a MSP - massive WD system, and PSR~J0514$-$4002E could also be a MSP - black hole system.}.
\citet{Ferdman:2020Natur} also derived the mass ratio of the system as $q=m_{c}/m_{p}=0.78\pm0.03$, making PSR J1913+1102 the most asymmetric DNS known to merge within a Hubble time.
The significant mass asymmetry in PSR J1913+1102 makes it a promising candidate for testing scalar-tensor theories of gravity \citep{Zhao:2022CQGra}. Its pronounced mass asymmetry, together with its spin and orbital properties, also provides a chance for probing DNS evolution and constraining the second SN, including the associated kick and progenitor properties.

The remainder of the paper is organized as follows. In Section \ref{sec:obs}, we describe the observations and the processing of the resulting pulsar timing data; in Section \ref{sec:results}, we present the new timing results. In Section \ref{sec:mass_grav}, we provide the mass measurements of this system and describe and discuss the gravity tests in GR and the scalar-tensor theory. In Section \ref{sec:evo}, we analyze the evolution of this DNS system and constrain the kick magnitude and progenitor mass for the second-formed NS. Finally, in Section \ref{sec:con_pro}, we summarise the results and discuss the potential of continued timing of PSR J1913+1102 for improved tests of gravity theories.   

\section{Observations and data analysis}\label{sec:obs}
Previous timing observations of this system were carried out using multiple Arecibo backends \citep{Ferdman:2020Natur}. 
At L band, observations were carried out with the Mock spectrometer \citep{Lazarus:2015ApJ} in incoherent mode at 1300 and 1450 MHz, and with the PUPPI\footnote{PUPPI is a clone of the Green Bank Ultimate Pulsar Processing Instrument \citep{DuPlain:2008}.} backend in coherent mode at 1400 MHz.
At S band, observations were carried out at 2350 MHz using PUPPI in coherent mode. Here, ``coherent'' and ``incoherent'' refer to the methods used to correct for dispersive delays caused by radio wave propagation through the interstellar medium (ISM), with coherent dedispersion providing a more complete correction by removing the dispersive response from the raw voltage data\footnote{The only disadvantage of this method is the much higher computational cost, which only became tractable for large bandwidths in the mid-2000s.}. After Arecibo was decommissioned, FAST, the most sensitive radio telescope at L band \citep{Jiang:2020RAA}, has become the primary instrument for observing this system. In this work, we present nearly 13-year timing results using both Arecibo and FAST.

\subsection{Data processing of FAST observations}
With FAST, we performed 27 epochs of observations with the central beam of the 19-beam receiver in incoherent mode \citep{Jiang:2020RAA}, with frequencies ranging from 1000 to 1500 MHz and 4096 frequency channels. We ultimately used only the band between 1020 to 1480 MHz because of the decrease in sensitivity at the edges of the band. 
Full-Stokes data were recorded, and each observation was preceded by a 1-minute calibration signal from a noise diode in the receiver for polarization calibration.
The observations were recorded with a sampling time of $49.152\,{\rm \upmu s}$ and digitised with 8 bits per sample.

\begin{figure}[htbp]
    \centering
    \includegraphics[width=0.49\textwidth]{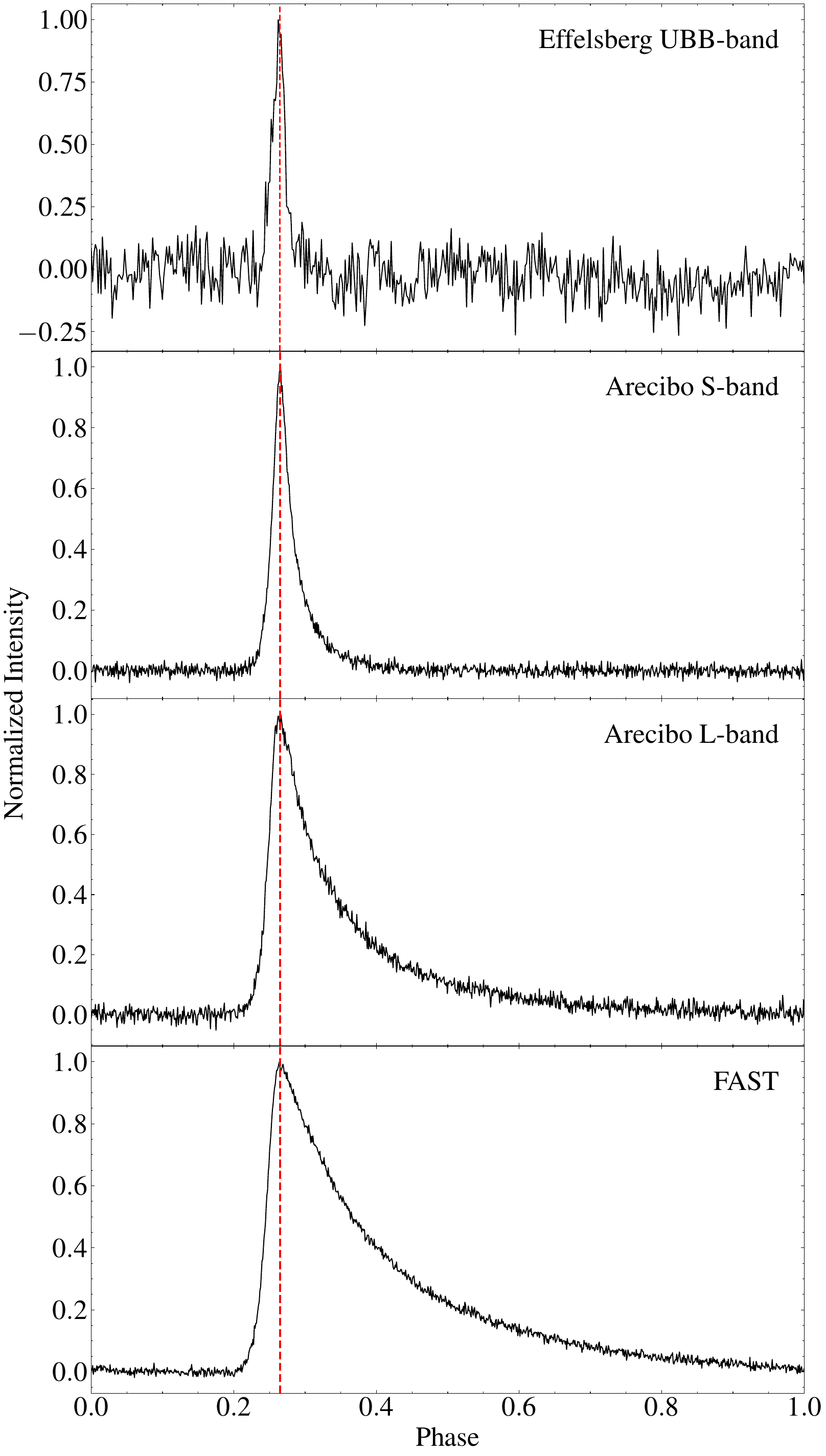}
    \caption{Frequency-averaged template profiles used in this work include Arecibo observations at S-band (central frequency 2350 MHz, bandwidth 800 MHz) and L-band (central frequency 1400 MHz, bandwidth 800 MHz), as well as FAST observations at L-band (central frequency 1250 MHz, bandwidth 480 MHz). For comparison, we also show the pulse profile obtained from a single sub-frequency band with a bandwidth of 750 MHz from the Effelsberg UBB receiver (central frequency 3350 MHz). \label{fig:profile}}
\end{figure}

We used \texttt{DSPSR}\footnote{\url{http://dspsr.sourceforge.net}} \citep{Straten:2011} to fold the data and derived the ToAs with \texttt{PSRCHIVE}\footnote{\url{http://psrchive.sourceforge.net/}} \citep{Hotan:2004,Straten:2012} software package. To construct a high signal-to-noise ratio (S/N) template, we integrated the 27 observations into a combined profile, which is shown in the bottom panel of Fig.\,\ref{fig:profile}. 
For comparison, the Arecibo S-band and L-band profiles, whose ToAs are used in this paper, are also displayed in the same figure. 
All of these profiles are broadened by interstellar scattering, with the effect being more severe at lower frequencies (see Fig.\,\ref{fig:profile}). 
As a consequence, the resulting ToAs of FAST are less precise than those obtained at Arecibo's higher-frequency S band, despite the higher sensitivity of FAST.
Observations at higher radio frequencies will thus be necessary to mitigate pulse broadening and significantly improve the timing precision of PSR~J1913+1102.

To further quantify the benefit of higher-frequency observations in reducing interstellar scattering, we performed an additional 8-hour observation using the Ultra Broad Band (UBB) receiver on the Effelsberg Telescope. This receiver is sensitive to frequencies between 1.2 and 6 GHz. We select the 2975-3725 MHz sub-band to display the pulse profile (see the top panel of Fig.\,\ref{fig:profile}). At these frequencies, the negligible scattering improves the timing precision of PSR~J1913+1102, despite the lower flux density.

For the FAST data, we employed the \texttt{paas} plugin in \texttt{PSRCHIVE} to perform a multi-Gaussian decomposition of the integrated profile. This procedure yielded a high-fidelity, noise-reduced template that minimizes systematic biases in the timing analysis.
The ToAs were subsequently obtained using the \texttt{pat} routine, which cross-correlates the template profile with each integrated data profile in the Fourier domain to determine a phase shift. This shift is then added to the observatory time stamp corresponding to each integrated profile to calculate the ToAs. Each ToA represents a four-minute integrated profile, providing a compromise between time resolution and S/N. The full observing band was subdivided into 4 frequency sub-bands to allow for fitting possible dispersion measure (DM) variations. All ToAs were created in the "princeton\footnote{\url{https://tempo.sourceforge.net/ref_man_sections/usage.txt}}" format of \texttt{TEMPO}.

\subsection{Timing procedure and DM variation modeling}

After obtaining the FAST ToAs, we combined them with the Arecibo ToAs used in \citet{Ferdman:2020Natur} in a joint timing analysis. To align profiles from different telescopes, backends, and frequency bands, we included JUMPs in the timing model to account for systematic offsets between data sets.

A phase-connected timing solution was obtained using \texttt{TEMPO}\footnote{\url{http://tempo.sourceforge.net/}} \citep{nds+15} with the DDGR and DDFWHE binary models, separately. The DDGR model \citep{Taylor:1989} is similar to the theory-independent Damour–Deruelle (DD) model \citep{DD:1986} but assumes that GR is the correct theory of gravity. In this model, instead of fitting for the PK parameters, one fits for the total system mass $m_{\rm tot}$ and the companion mass $m_{\rm c}$ directly; all relativistic effects are calculated internally based on the predictions of GR.
The second model, DDFWHE (implemented in \texttt{TEMPO} by \citealt{Weisberg:2016ApJ}), is also derived from the DD model and is theory-independent. The only difference lies in the parameterisation of the Shapiro delay:
instead of the Shapiro ``range'' ($r$) and ``shape'' ($s$) parameters, we fit for the othometric amplitude ($h_{3}$) and ratio ($\varsigma$) parameters \citep{Freire:2010MNRAS}; the reasons for this adoption are explained below (see Sec.\,\ref{sec:shapiro}).

\begin{figure}[htbp]
    \centering
    \includegraphics[width=0.49\textwidth]{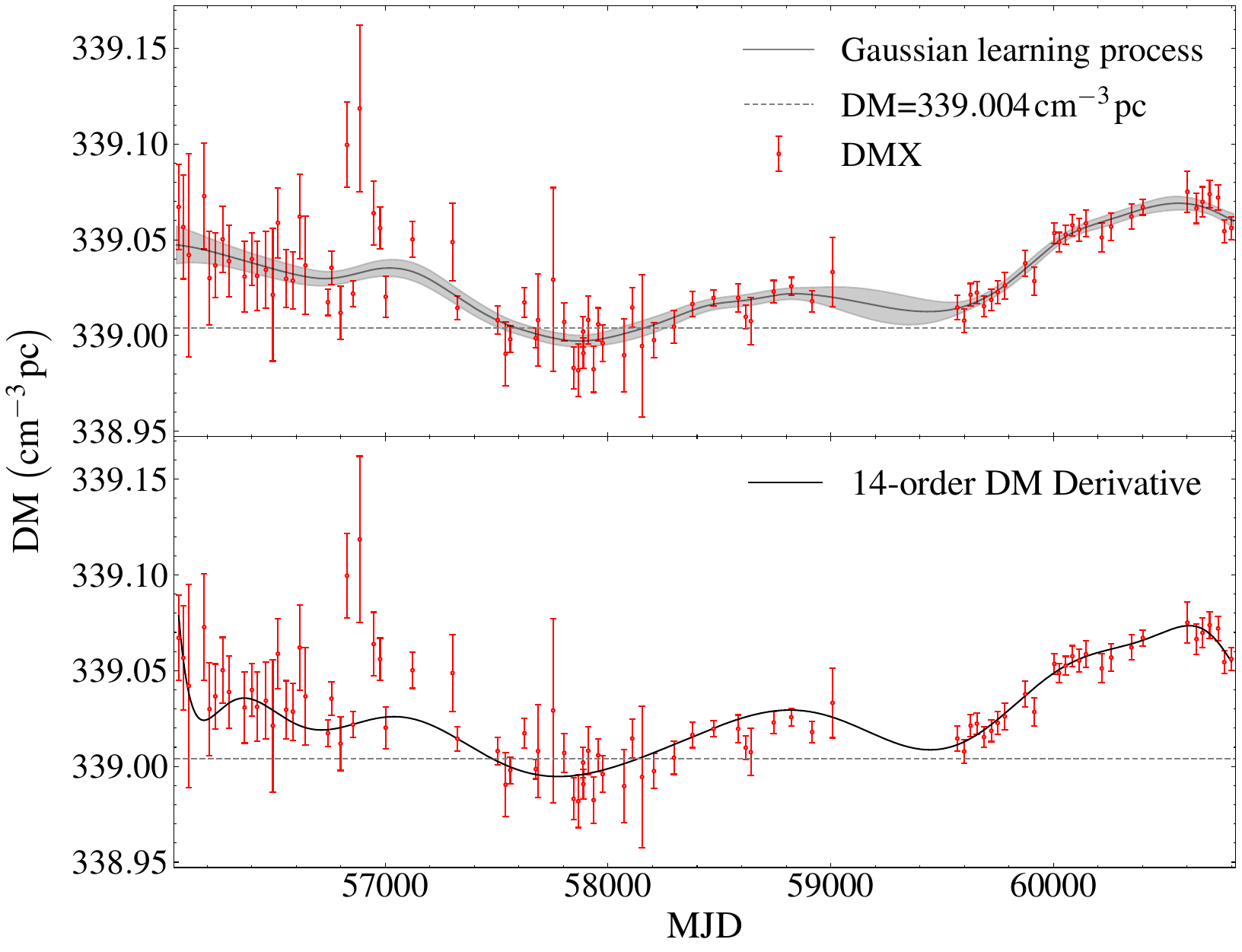}
    \caption{DM variations and fit results for PSR J1913+1102. In both panels, the red dots represent the DMX (20-day interval) values obtained from the DDFWHE model. The top panel shows the Gaussian-process fit (gray solid line) with the shaded region indicating the 1-$\sigma$ uncertainty of the fitted offsets, while the bottom panel displays a 14-order DM-derivative fit (black solid line).\label{fig:dmvariation}}
\end{figure}

PSR J1913+1102 has a very low Galactic latitude ($b = 0.19^{\circ}$) and a large DM of $339 \, \rm cm^{-3} \,pc$. This means that the radio waves travel through a dense, complex ISM within the Galactic plane. Consequently, as the system and the Earth move, there will also be large, complex and unpredictable temporal variations in the electron column density probed by the changing line of sight. In addition, intrinsic temporal fluctuations within the ISM may also contribute to the observed DM variability. 
This can be seen within Fig.\,\ref{fig:dmvariation}, in which we plot short-term (20 days) DM measurements over time; these were obtained using the DMX model\footnote{The DMX model accounts for temporal variations in DM by grouping ToAs into discrete epochs and fitting an independent DM value for each epoch.} \citep{Demorest:2013ApJ} within the DDFWHE binary framework.

Properly modeling such variations is essential, since uncorrected changes can bias the measurement of other timing parameters, particularly the pulsar's proper motion. In this work, we modeled the DM variations using a Gaussian learning process\footnote{This method models the DM as a correlated stochastic process, allowing its temporal evolution to be inferred directly from the data without assuming a prescribed functional form.} implemented in the \texttt{Python} library \texttt{George} \citep{Rasmussen:2006gpml}, and adopted the \texttt{Matern32Kernel}.
The top panel of Fig.\,\ref{fig:dmvariation} shows the DM variation fitted with this method\footnote{We have also generated different series of DM measurements, using intervals of 10 and 40 days for data averaging, applied a Gaussian process to these datasets, and then used the resulting DM models in our timing data set. In all cases, we obtained consistent timing results, which means that, unlike in the timing solutions obtained directly with the DMX model, the timing parameters do not depend on the time intervals used to average the data.}. The gray solid line represents the central values of the DM offsets, while the gray shaded region indicates the 1-$\sigma$ uncertainty. 
After obtaining the DM offsets using a Gaussian process, we write them into the last column of these TOA lines.
We then set NDDM = 1 in the ephemeris of the pulsar to apply these DM offsets during the pulsar timing fit; this only works if the ToAs are written in the ``princeton'' format.

We verified these results by fitting the temporal DM variations with a model that contains 14 time derivatives of the DM. The bottom panel of Fig.\,\ref{fig:dmvariation} presents the fitted curve using this model (black solid line), which closely follows the DM variations from the DMX model. 
The justification of this model, and the resulting timing parameters, are discussed in detail in Appendix \ref{sec:appendix}. The difference between the parameters of both models is statistically insignificant.

\section{Results}\label{sec:results}
\begin{figure*}[htb]
  \centering

  \begin{subfigure}[b]{\linewidth}
    \centering
    \includegraphics[width=0.99\linewidth]{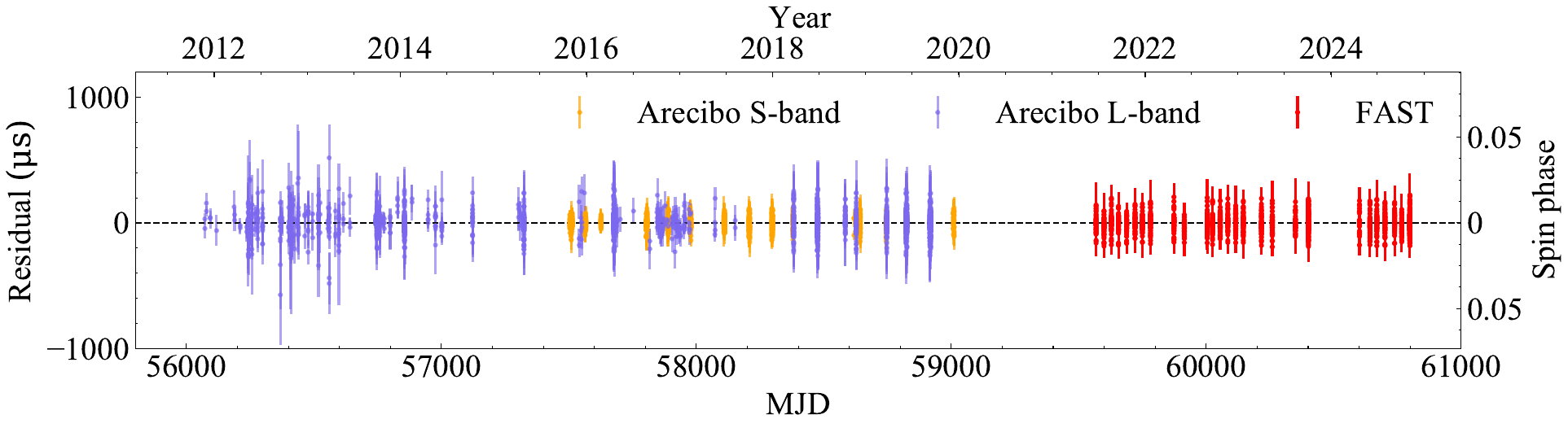}
    \label{fig:1a}
  \end{subfigure}

  \vskip\baselineskip 

  \begin{subfigure}[b]{\linewidth}
    \centering
    \includegraphics[width=0.99\linewidth]{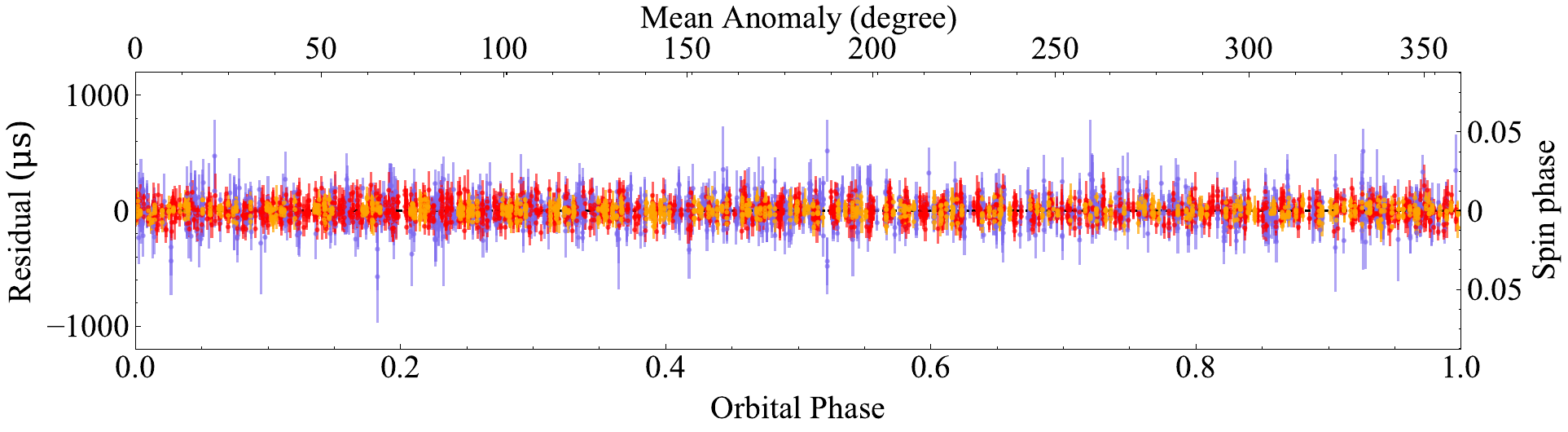}
    \label{fig:1b}
  \end{subfigure}

  \caption{Residuals (ToAs minus DDFWHE model predictions) from timing analysis of PSR J1913+1102 are shown with 1-$\sigma$ uncertainties. The residuals in orange and medium slate blue represent the Arecibo S-band and L-band data, respectively, and the red points are obtained from FAST. The top figure shows the residuals as a function of time. The bottom figure shows the residuals as a function of the orbital phase. The ToA residuals are consistent with their uncertainties, and no unmodeled trends are evident, indicating that the DDFWHE timing solution adequately describes the system within the measurement precision.}
  \label{fig:residuals}
\end{figure*}

The timing results derived from different binary models are summarized in Tab.\,\ref{tab:timing_gp}, with all uncertainties quoted at the 1-$\sigma$ confidence level. The residuals (the difference between the ToAs and the predictions of the DDFWHE model for the same rotation) are shown in Fig.\,\ref{fig:residuals}.
For the 5713 ToAs used in our timing fit, we obtain a weighted rms residual of $49.7\,{\rm \upmu s}$ and a reduced $\chi^{2}$ of 1.0029.
As shown in this figure, these residuals are consistent with zero within their uncertainties and have no visible deviating trends, both as  functions of MJD and orbital phase. This behavior demonstrates that the DDFWHE model provides a good description of the timing data.
In Fig.\,\ref{fig:residuals}, the residuals obtained from FAST ToAs are shown in red, while those from Arecibo are shown in orange and blue for the S-band and L-band data, respectively. 
Although FAST has superior sensitivity compared to Arecibo at L band, its 1-1.5 GHz data do not yield a significant improvement in ToA precision.
As discussed in Sec.\,\ref{sec:obs}, this reflects the frequency dependence of scatter broadening, with stronger effects at lower frequencies (see Fig.\,\ref{fig:profile}).

\begin{figure}[htbp]
    \centering    \includegraphics[width=0.499\textwidth]{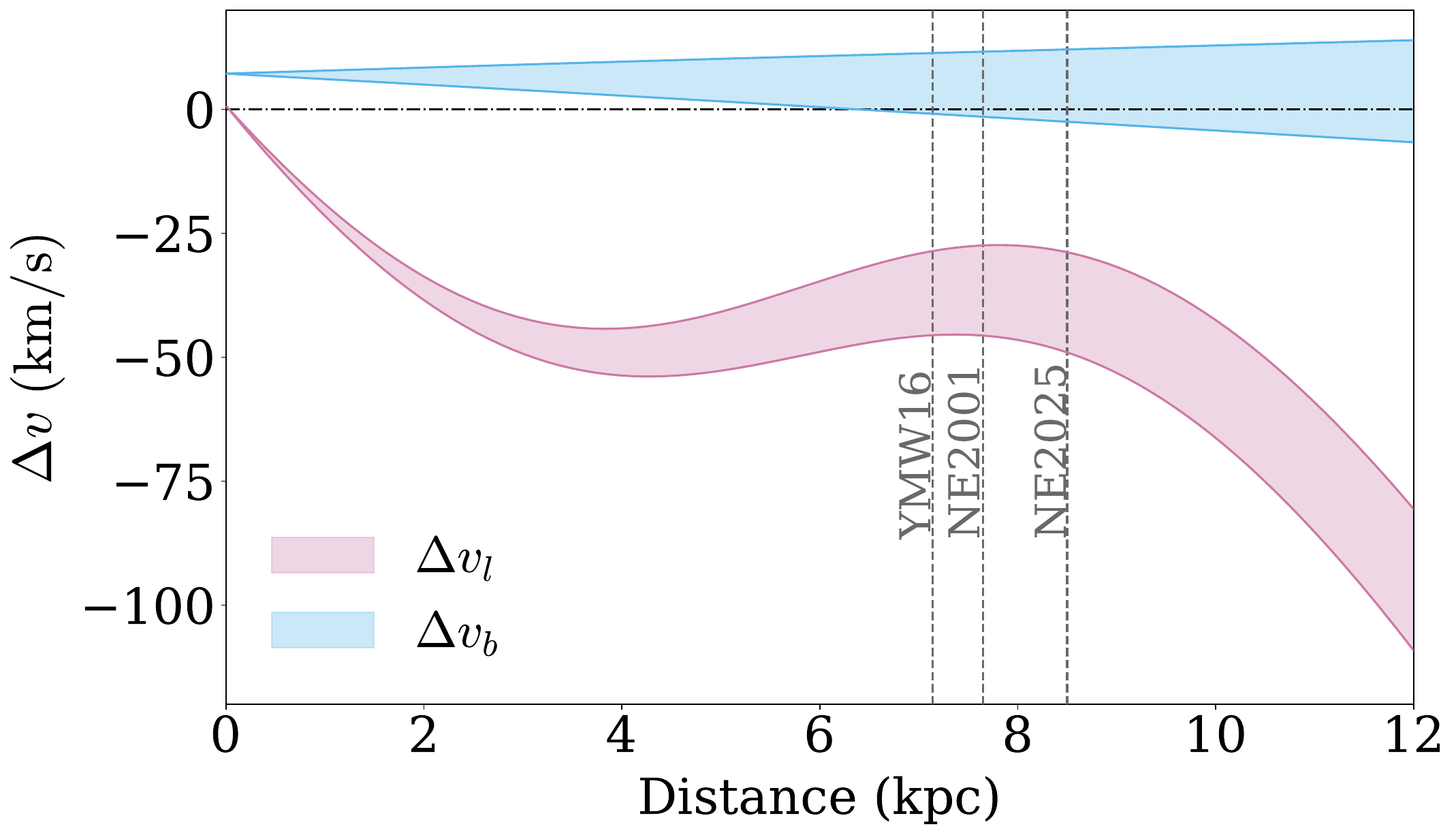}
    \caption{Two components of the transverse peculiar velocity (w.r.t. LCF) of PSR~J1913+1102 along Galactic longitude ($\Delta v_l$) and latitude ($\Delta v_b$), calculated assuming distances from the Solar System ranging from 0 to 12\,kpc (see text for detailed description). The shaded regions between them indicate the 1-$\sigma$ uncertainties. The three vertical gray dashed lines mark the distances inferred from the YMW16, NE2001 and NE2025 models.\label{fig:v_diff}}
\end{figure}

\subsection{Astrometric parameters}
The measured position coordinates of this pulsar are a right ascension ($\alpha$, J2000) of $19^\mathrm{h}$ $13^\mathrm{m}$ 29\fs05371(4) and a declination ($\delta$, J2000) of $11^\circ$ $02\arcmin$ $05\farcs7020(11)$.
Transforming the position into Galactic coordinates gives a longitude of $l=45.25^{\circ}$ and a latitude of $b=0.19^{\circ}$, placing the pulsar essentially within the Galactic plane. Combining this with the measured DM ($339 \, \rm cm^{-3} \,pc$), we estimate its distance from the Solar System ($d$) using Galactic electron-density models. The NE2001 model \citep{Cordes:ne2001} gives $d_{\rm DM}=7.65\,$kpc, while the YMW16 model \citep{Yao:ymw2016} yields $d_{\rm DM}=7.14\,$kpc. We include the updated Galactic electron-density model NE2025 \citep{ne2025} in our distance estimate, yielding $d_{\rm DM} = 8.50\,$kpc\footnote{We use the Fortran version of NE2025 to get this estimated distance.}. All of these provide mutually consistent results within the relative 20\% uncertainty estimated by \cite{Cordes:ne2001}.

Combining 13 years of ToAs from FAST and Arecibo, we obtain a new proper motion measurement of $\mu_{\alpha}=-3.52(15)\,{\rm mas\,yr^{-1}}$, $\mu_{\delta}=-6.85(27)\,{\rm mas\,yr^{-1}}$. These values are respectively 3.3 and 3.7 times more precise than the measurement by \citet{Ferdman:2020Natur}.
Normally, the uncertainty of the proper motion improves with $T^{-3/2}$, where $T$ is the timing baseline. Our timing baseline is $\sim 13/8$ times larger than that reported by \cite{Ferdman:2020Natur}, so from this scaling, we would expect an improvement in the precision by a factor of $\sim 2$. Therefore, the extended data span does not fully explain the improvement; additional factors are the higher cadence of the FAST data, the improved DM modeling, and the decrease in the correlation with other parameters such as the spin frequency and its derivative. This conclusion also applies to other measured parameters, although their dependence on the timing baseline differs \citep{Damour:1992PrD}.

The total proper motion, $\mu = \sqrt{\mu_{\alpha}^2 + \mu_{\delta}^2} = 7.71(25)\,{\rm mas\,yr^{-1}}$, is 1.5 $\sigma$ smaller than the value reported by \citet{Ferdman:2020Natur}, $\mu = 9.2(1.0)\,{\rm mas\,yr^{-1}}$.
From the timing solution listed in Tab.\,\ref{tab:timing_gp}, using the distance $d_{\rm DM}=7.65\,$kpc from the NE2001 model with $20\%$ Gaussian uncertainty,
we derive a heliocentric transverse velocity, $v_{\rm T}^{\rm SSB} = 4.74\,{\mathrm{km\,s^{-1}}\left(\frac{\mu}{\mathrm{mas\,yr^{-1}}}\right)\left(\frac{d}{\rm kpc}\right)}=279(56)\,{\rm km\,s^{-1}}$.

By transforming the measured proper motions $\mu_{\alpha}$ and $\mu_{\delta}$ in Galactic coordinates we obtain $\mu_{l}=-7.71(25)\,{\rm mas\,yr^{-1}}$ and $\mu_{b}=-0.05(18)\,{\rm mas\,yr^{-1}}$, corresponding to a Galactic position angle of the proper motion, $\tan^{-1}\left({\mu_{l}}/{\mu_{b}}\right)=269.6(1.3)^{\circ}$. This indicates a predominantly westward motion in the Galaxy with a negligible vertical component.

Using the \texttt{MWPotential2014} model \citep{MWPotential2014}, we calculated the expected proper motions for points located at the position of PSR J1913+1102 (with distances given by $d \in    [0,\,12]\,{\rm kpc}$) that are co-rotating with the Galaxy, i.e., they are in the local co-rotating frame (LCF). Subtracting these from the observed pulsar proper motion yields a (distance-dependent) proper motion difference. Multiplying the latter by the corresponding distance results in  the {\em transverse peculiar velocity}. In Fig.\,\ref{fig:v_diff}, we plot each component of the transverse peculiar velocity against distance from the Solar System.
The vertical component of the peculiar velocity, $\Delta v_{b}$, corresponds to a nearly constant northward velocity offset of about $+10\,{\rm km\,s^{-1}}$. Along the $l$ direction, the peculiar velocity $\Delta v_{l}$ is approximately $-40\,{\rm km\,s^{-1}}$ at the distances indicated by the three Galactic electron-density models, indicating a westward motion with a weak dependence on the distance over the range of 4 to 10\,kpc. This suggests that a relatively mild kick was imparted onto the second NS by the SN event that formed it.

\begingroup
\renewcommand{\arraystretch}{0.93}
\begin{table*}
\begin{center}
  \caption{{Fitted and derived parameters for PSR J1913+1102. Numbers in parentheses indicate the 1-$\sigma$ uncertainties in the last quoted digits, as returned by \texttt{TEMPO} and scaled to yield a reduced $\chi^2 = 1$. Values in square brackets are derived from the DDGR model under the assumption that GR is correct.}\label{tab:timing_gp}}
  \begin{footnotesize}
  \begin{tabular}{lcc}
  \hline\hline
  \multicolumn{3}{c}{Data and data reduction parameters} \\
  \hline
  Solar System ephemeris \dotfill & \multicolumn{2}{c}{DE440} \\
  Time Units \dotfill & \multicolumn{2}{c}{TDB} \\
  Clock \dotfill & \multicolumn{2}{c}{TT(BIPM2023)} \\
  Epoch (MJD) \dotfill & \multicolumn{2}{c}{57504.531453} \\
  Span of Timing Data (MJD) \dotfill & \multicolumn{2}{c}{56072--60799} \\
  RMS Residual ($\rm \upmu s$) \dotfill & 49.705 & 49.715 \\
  $\chi^2$ \dotfill & 5707.31 & 5707.54 \\
  Reduced $\chi^2$ \dotfill & 1.002866 & 1.003081\\
  \hline\hline
  \multicolumn{3}{c}{Measured parameters} \\
  \hline
  Right Ascension, $\alpha$ (J2000.0) (h:m:s) \dotfill & \multicolumn{2}{c}{19:13:29.05371(4)} \\
  Declination, $\delta$ (J2000.0) ($\deg$:$\arcmin$:$\arcsec$) \dotfill & \multicolumn{2}{c}{11:02:05.7020(11)} \\
  Proper motion in Right Ascension, $\mu_\alpha$ ($\rm mas\,yr^{-1}$) \dotfill & \multicolumn{2}{c}{$-3.52(15)$} \\
  Proper motion in Declination, $\mu_\delta$ ($\rm mas\,yr^{-1}$) \dotfill & \multicolumn{2}{c}{$-6.85(27)$} \\
  Pulse frequency, $\nu$ (Hz) \dotfill & \multicolumn{2}{c}{36.6501648629335(10)} \\
  First derivative of pulse frequency, $\dot \nu$ (Hz $\mathrm{s^{-1}}$) \dotfill & \multicolumn{2}{c}{$-2.1014(3)\times 10^{-16}$} \\
  Second derivative of pulse frequency, $\ddot \nu$ (Hz $\mathrm{s^{-2}}$) \dotfill & \multicolumn{2}{c}{$-2.71(23)\times 10^{-27}$} \\
  Dispersion measure, DM ($\mathrm{cm^{-3}\,pc}$) \dotfill & \multicolumn{2}{c}{339.0034(19)} \\
  Binary model \dotfill & DDFWHE & DDGR \\
  Orbital period, $P_\mathrm{b}$ (d) \dotfill & 0.20625233485(8) &  0.20625233492(7) \\
  Projected semimajor axis, $x$ (s) \dotfill & 1.754620(8) & 1.7546324(20) \\
  Orbital eccentricity, $e$ \dotfill & 0.0895303(21) & 0.0895276(11) \\
  Epoch of periastron, $T_\mathrm{0}$ (MJD) \dotfill & 57504.5314518(5) & 57504.5314521(5) \\
  Longitude of periastron, $\omega$ (deg) \dotfill & 283.7873(10) & 283.788(9) \\
  Rate of periastron advance, $\dot \omega$ (deg\,$\rm yr^{-1}$) \dotfill & 5.65112(26) &  [5.6513344] \\
  Change in orbital period, $\dot P_\mathrm{b}$ ($10^{-12}\, \mathrm{s\;s^{-1}}$) \dotfill & $-0.455(6)$ & [$-$0.4621390] \\
  Einstein delay amplitude, $\gamma$ (s) \dotfill & 0.000481(4) & [0.00048104] \\
  Total system mass, $m_{\rm tot}$ ($\mathrm{M_{\odot}}$) \dotfill & - & 2.88965(17) \\
  Companion mass, $m_{\rm c}$ ($\mathrm{M_{\odot}}$) \dotfill & - & 1.291(8) \\
  Orthometric ratio of Shapiro delay, $\varsigma$ \dotfill & 0.509$^{\rm a}$ & - \\
  Orthometric amplitude of Shapiro delay, $h_3$ ($\upmu$s) \dotfill & 1.7(6) & - \\
  \hline\hline
  \multicolumn{3}{c}{Derived parameters} \\
  \hline
  Galactic longitude, $l$ (deg) \dotfill & \multicolumn{2}{c}{45.25} \\
  Galactic latitude, $b$ (deg) \dotfill & \multicolumn{2}{c}{0.19} \\
  DM-derived distance (NE2001), $d_\mathrm{DM}$ (kpc) \dotfill & \multicolumn{2}{c}{7.65} \\
  DM-derived distance (NE2025), $d_\mathrm{DM}$ (kpc) \dotfill & \multicolumn{2}{c}{8.50} \\
  DM-derived distance (YMW16), $d_\mathrm{DM}$ (kpc) \dotfill & \multicolumn{2}{c}{7.14} \\
  Galactic height (NE2001), $z_\mathrm{DM}$ (kpc) \dotfill & \multicolumn{2}{c}{0.026} \\
  Galactic height (NE2025), $z_\mathrm{DM}$ (kpc) \dotfill & \multicolumn{2}{c}{0.028} \\
  Galactic height (YMW16), $z_\mathrm{DM}$ (kpc) \dotfill & \multicolumn{2}{c}{0.024} \\
  Magnitude of proper motion, $\mu$ ($\rm mas \, yr^{-1}$) \dotfill & \multicolumn{2}{c}{7.71(25)} \\
  Position angle of proper motion, $\Theta_{\mu}$ ($\deg$, J2000) \dotfill & \multicolumn{2}{c}{207.2(1.3)} \\
  Position angle of proper motion, $\Theta_{\mu}$ ($\deg$, Galactic) \dotfill & \multicolumn{2}{c}{269.6(1.3)} \\
  Heliocentric transverse velocity, $v_{\rm T}^{\rm SSB}$ ($\rm km \, s^{-1}$) \dotfill & \multicolumn{2}{c}{279(56)$^{\rm b}$} \\
  Spin period, $P$ (s) \dotfill & \multicolumn{2}{c}{0.027285006868042(7)} \\
  Observed spin period derivative, $\dot P$ ($\rm s\, s^{-1}$) \dotfill & \multicolumn{2}{c}{1.56444(22)$\times 10^{-19}$} \\
  Intrinsic spin period derivative, $\dot P^\mathrm{int}$ ($\rm s\, s^{-1}$) \dotfill & \multicolumn{2}{c}{1.483(23)$\times 10^{-19}$} \\
  Characteristic age, $\tau_\mathrm{c} = P/2 \dot{P}$ (Gyr) \dotfill & \multicolumn{2}{c}{2.91(4)} \\
  Surface magnetic field, $B_\mathrm{S} = 3.2 \times 10^{19}\sqrt{ P \dot P}$ ($10^{9}$G) \dotfill & \multicolumn{2}{c}{2.03(2)} \\
  Spin-down luminosity  $\dot{E}$ ($10^{32}$ erg/s)\dotfill  & \multicolumn{2}{c}{2.88(4)} \\
  Merger time (Myr)\dotfill  & \multicolumn{2}{c}{470} \\
  Mass function, $f_\mathrm{mass}$ ($\mathrm{M_{\odot}}$) \dotfill & 0.1363434(20) & 0.1363463(5) \\
  Pulsar mass, $m_{\rm p}$ ($\mathrm{M_{\odot}}$) \dotfill & 1.599(8)  & 1.599(8) \\
  Total system mass, $m_{\rm tot}$ ($\mathrm{M_{\odot}}$) \dotfill & 2.88948(20) & - \\
  Sine of orbital inclination, $\sin(i)$ \dotfill & 0.809 & 0.811(5) \\
  Orbital inclination, $i$\,(deg) \dotfill & $54$ &$54.2(5)$ \\
  Companion mass, $m_{\rm c}$ ($\mathrm{M_{\odot}}$) \dotfill & 1.290(8) & - \\
  \hline
  \multicolumn{3}{l}{a: $\varsigma$ is a fixed value in DDFWHE model.}\\
  \multicolumn{3}{l}{b: Used NE2001 model to get distance with $20\%$ uncertainty in a normal distribution.}
\end{tabular} 
\end{footnotesize}
\end{center}
\end{table*}   
\endgroup

\renewcommand{\arraystretch}{1.34}
\begin{table}
\centering
\caption{Relative Galactic acceleration difference ($\dot{P}_{\rm b}^{\rm Gal}$) and Shklovskii effect ($\dot{P}_{\rm b}^{\rm Shk}$) contributions to $\dot{P}_{\rm b}^{\rm obs}$ are estimated using three distances from NE2001, YMW16, and NE2025 and assumed a $20\%$ uncertainty. We also present the total non-intrinsic contributions ($\dot{P}_{\rm b}^{\rm ext}$), the intrinsic contributions ($\dot{P}_{\rm b}^{\rm intr}$) and the GR prediction for the orbital decay ($\dot{P}_{\rm b}^{\rm GR}$). The uncertainties in these results are derived from MC simulations.
\label{tab:extralpbdot_dmgp}}
\begin{tabular}{crrr}
\hline\hline
~ & NE2001 & YMW16 & NE2025 \\
\hline
$\dot{P}_{\rm b}^{\rm obs}\,({\rm fs\,s^{-1}})$ & \multicolumn{3}{c}{$-454.8^{+5.7}_{-5.7}$} \\
$\dot{P}_{\rm b}^{\rm Gal}\,({\rm fs\,s^{-1}})$ & $-15.0^{+4.6}_{-2.8}$ & $-13.7^{+4.6}_{-3.3}$ &{$-16.8^{+4.2}_{-1.8}$} \\
$\dot{P}_{\rm b}^{\rm Shk}\,({\rm fs\,s^{-1}})$ & $~19.6^{+4.2}_{-4.1}$ & $~~18.3^{+3.9}_{-3.8}$ &{$21.8^{+4.6}_{-4.5}$}\\
$\dot{P}_{\rm b}^{\rm ext}\,({\rm fs\,s^{-1}})$ & $5.3^{+1.6}_{-1.5}$&  $5.2^{+1.4}_{-1.5}$ &{$5.6^{+2.5}_{-1.7}$} \\
$\dot{P}_{\rm b}^{\rm intr}\,({\rm fs\,s^{-1}})$ & $-460.2^{+6.0}_{-6.0}$ & $-460.1 ^{+5.9}_{-5.9}$ & $-460.8^{+6.1}_{-6.2}$ \\
$\dot{P}_{\rm b}^{\rm GR}\,({\rm fs\,s^{-1}})$ & \multicolumn{3}{c}{$-462.0^{+0.5}_{-0.5}$} \\
\hline
\end{tabular}
\end{table}

\subsection{Rate of periastron advance}\label{sec:omegadot}
Due to the effects of general relativity, the longitude of the periastron ($\omega$) gradually increases with time; this is quantified by the PK parameter known as the rate of periastron advance, $\dot{\omega}$.
The updated timing results yield a measurement of $\dot{\omega} = 5.65112(26)\,{\rm deg\,yr^{-1}}$ with the DDFWHE model. This value is higher than the value reported by \citet{Ferdman:2020Natur} by 1.4 $\sigma$, but is measured with twice the precision.
In this system, the contribution of the Kopeikin term \citep{Kopeikin:1996}, arising from the proper motion of the binary, is at most $\sim2\times10^{-6}\,{\rm deg\,yr^{-1}}$, and is therefore negligible. Thus, under the assumption that the effect is purely relativistic, the observed $\dot{\omega}$ can be used to derive the total mass of the binary system:
\begin{equation}
    m_{\rm tot} = \frac{1}{T_{\odot}}\left[\frac{\dot{\omega}(1-e^{2})}{3}\right]^{3/2}\left(\frac{P_{\rm b}}{2\pi}\right)^{5/2}=2.88948(20)\,{\rm M_{\odot}}\,,
\end{equation}
where $T_{\odot}\equiv({\mathcal{G M})_{\odot}^{\rm N}/c^{3}}=4.925490947641266978...\,{\upmu \rm s}$ is the solar mass parameter \citep[$(\mathcal{G M})_{\odot}^{\rm N}=1.3271244\times10^{20}\,{\rm m^{3}\,s^{-2}}$,][]{Prsa:2016} in time units, $c$ is the speed of light, and $e$ is the orbital eccentricity.\footnote{In equations where $T_{\odot}$ appears the mass values are adimensional; however we add the symbol $\rm M_{\odot}$ to make it clear that these indicate multiples of the solar mass parameter, $m_i \leftrightarrow G m_i / (\mathcal{G M})_{\odot}^{\rm N}$.} Our estimate of $m_{\rm tot}$ is 1.3 $\sigma$ larger than the previously published value  \citep{Ferdman:2020Natur}, but three times more precise.

\subsection{Einstein delay}\label{sec:gamma}
To leading order, the Einstein delay can be understood as the combination of the variation with orbital phase of two time dilation effects: the second-order Doppler effect and the gravitational redshift. This variation is proportional to the orbital eccentricity and both delays add in phase with each other, with both increasing fastest at periastron. The amplitude of the Einstein delay is quantified by the PK parameter $\gamma$, which in GR is given by
\begin{equation}
    \gamma=T_{\odot}^{2/3}\left(\frac{P_{\rm b}}{2\pi}\right)^{1/3}e\frac{\left(2m_{\rm c}+m_{\rm p}\right)m_{\rm c}}{(m_{\rm p}+m_{\rm c})^{4/3}}\,. \label{eq:gamma}
\end{equation}
In this work, we measure $\gamma=0.000481(4)\,$s, which is 1-$\sigma$ consistent with the previously published value but with a 3.7-fold improvement in precision.
Using the total mass derived from $\dot{\omega}$ together with the measured $\gamma$, we determine the individual component masses via Eq.\,(\ref{eq:gamma}), to be $m_{\rm p} = 1.599(8)\,{\rm M_{\odot}}$ and $m_{\rm c} = 1.290(8)\,{\rm M_{\odot}}$, respectively. These component mass measurements agree with those reported by \citet{Ferdman:2020Natur}, but at a substantially higher precision, improved by a factor of 3.7.
The newly measured mass ratio is $q = 0.807(8)$, confirming that this system remains the most asymmetric DNS among known merging systems.

\subsection{Shapiro delay}\label{sec:shapiro}
The Shapiro delay in a binary system is a relativistic signal propagation delay caused by the curvature of spacetime near the companion \citep{Shapiro:1964}. \cite{Blandford:1976}, and later
\cite{DD:1986} quantify this effect in binary pulsars to leading order using the PK parameters known as ``range'' ($r$) and ``shape'' ($s$); in GR, $s = \sin i$ and $r = T_\odot m_\mathrm{c}$. However, in low-eccentricity systems, particularly when the orbital inclination angle $i$ is far from edge-on, $r$ and $s$ become strongly correlated, leading to a low numerical precision.
To overcome this limitation, \citet{Freire:2010MNRAS} introduced an orthometric parametrization with two new PK parameters, the orthometric amplitude $h_{3}$ and the orthometric ratio $\varsigma$, which can be expressed as a function $r$ and $s$ as \citep{Freire:2010MNRAS}, 
\begin{align}
    \varsigma &=\frac{s}{1+\sqrt{1-s^2}}\,, \\
    \quad h_3 &=r \varsigma^3\,.
    \label{eq:shapiro}
\end{align}
The use of the orthometric parameters significantly reduces correlations and thereby improves the description of the localisation of the system in the mass-mass diagram; the $h_3$ parameter in particular provides an improved test of gravity theories compared to the $r$ parameter \citep{Freire:2024,Meng:2025}.
This parametrisation was implemented as the DDFWHE orbital model \citep{Weisberg:2016ApJ}, and is adopted in our analysis because the orbital inclination is far from edge-on \citep{Ferdman:2020Natur}.

In our ToA fit, we fixed the orthometric ratio at $\varsigma = 0.509$, derived from the $\sin i = 0.809$ of the best DDGR solution, and fitted for $h_{3}$, obtaining $h_{3} = 1.7(6)\,\upmu{\rm s}$. 
From the masses and $i$ derived from $\dot{\omega}$ and $\gamma$, we obtain from Eq.\,(\ref{eq:shapiro}) a prediction for $h_{3}$ of $0.84\,\upmu\rm s$; the measured value is 1.2 $\sigma$ larger than the GR prediction.
Owing to the low correlation between $h_3$ and
$\varsigma$, the assumption of the particular value of $\varsigma$ does not have much influence in the value of $h_3$.

\subsection{Orbital period decay rate \label{sec:pbdot}}
One of the main results that comes from our timing analysis is the observed variation of the orbital period, $\dot{P}_{\rm b}^{\rm obs} = -454.8(5.7)  \, \rm fs\,s^{-1}$.
This is 1-$\sigma$ consistent with the value measured by \cite{Ferdman:2020Natur} but five times more precise.

This variation is dominated by an intrinsic contribution arising from orbital decay driven by the emission of GWs. In GR, GWs are quadrupolar to leading order, and the orbital decay is given by \citep{Peters:1964}:
\begin{align}
  \dot{P}^{\rm GR}_{\rm b} &=-\frac{192 \pi}{5} T_{\odot}^{5 / 3}\left(\frac{P_{\rm b}}{2 \pi}\right)^{-5 / 3} f(e) \frac{m_{\rm p} m_{\rm c}}{\left(m_{\rm p}+m_{\rm c}\right)^{1 / 3}}\,, \label{eq:pbdot}
\end{align}
where
\begin{equation}
    f(e) \equiv \frac{1+(73/24)e^{2}+(37/96)e^{4}}{(1-e^{2})^{7/2}}\,. \label{eq:fe}
\end{equation}
However, kinematic contributions — such as the Shklovskii effect and the differential Galactic acceleration between the Solar System barycenter (SSB) and the pulsar system — also affect $\dot{P}_{\rm b}^{\rm obs}$:
\begin{equation}\label{eq:pbobs_eq}
    \left(\frac{\dot{P}_{\rm b}}{{P}_{\rm b}}\right)^{\rm intr}=\left(\frac{\dot{P}_{\rm b}}{{P}_{\rm b}}\right)^{\rm obs}-\left(\frac{\dot{P}_{\rm b}}{{P}_{\rm b}}\right)^{\rm Shk}-
    \left(\frac{\dot{P}_{\rm b}}{{P}_{\rm b}}\right)^{\rm Gal}\,.
\end{equation}
Hence, to estimate the intrinsic orbital decay rate, $\dot{P}_{\rm b}^{\rm intr}$, the contributions from these kinematic effects must be subtracted.
The Shklovskii effect \citep{Shklovskii:1970}, arising from the transverse motion of the pulsar system, depends on its proper motion and distance, and is given by
\begin{align}
   & \left(\frac{\dot{P}_{\mathrm{b}}}{P_{\mathrm{b}}}\right)^{\mathrm{Shk}}=\frac{\mu^2 d}{c}.
\end{align}
In this work, we estimate $d$ using the NE2001, YMW16, and NE2025 models, each with a $20\%$ uncertainty modeled as a Gaussian distribution.
Combining these distances with the proper motion obtained from timing, we calculate the Shklovskii contribution, as summarized in Tab.\,\ref{tab:extralpbdot_dmgp}.

The contribution from the relative Galactic acceleration difference is,
\begin{align}
   & \left(\frac{\dot{P}_{\mathrm{b}}}{P_{\mathrm{b}}}\right)^{\mathrm{Gal}}=-\frac{\left(\textbf{a}_{\rm PSR}-\textbf{a}_{\rm SSB}\right) \cdot \textbf{n}}{c}\,,\label{eq:galacc}
\end{align}
where $\textbf{n}$ is the unit vector pointing from the SSB to the pulsar, $\textbf{a}_{\rm PSR}$ is the acceleration of the pulsar, and $\textbf{a}_{\rm SSB}$ is the acceleration of the Solar System. Both accelerations are determined by the Galactic potential model adopted.
Multiple models have been proposed to describe the gravitational potential of the Milky Way. 
In this work, we employ \texttt{Galpy}\footnote{http://github.com/jobovy/galpy} \citep{Bovy:2015ApJS} to calculate the gravitational potential difference between the pulsar and the SSB, and then get accelerations in Eq.\,(\ref{eq:galacc}). 
The calculations are performed independently with two Galactic potential models, \texttt{MWPotential2014} \citep{MWPotential2014} and \texttt{McMillan17} \citep{McMillan:2017}, yielding results that are consistent within 1 $\sigma$ \footnote{Using the \texttt{McMillan17} Galactic potential and the distance estimated from the NE2001 model, we obtain $\dot{P}_{\rm b}^{\rm Gal}=-14.1^{+2.7}_{-1.9}\,{\rm fs\,s^{-1}}$. This demonstrates that the choice of Galactic potential model has only a minor impact on $\dot{P}_{\rm b}^{\rm Gal}$ for this system. Hence, the two Galactic potential models give $\dot{P}_{\rm b}^{\rm intr}$ values that are consistent within 1 $\sigma$, indicating that the choice of potential has a negligible effect on the result.}.

Tab.\,\ref{tab:extralpbdot_dmgp} summarizes the results obtained with the \texttt{MWPotential2014} model, using distances inferred from the three electron-density models under the same uncertainty assumptions as described above. 
The uncertainties of all input parameters are propagated through Monte Carlo (MC) simulations to determine final uncertainties.
The two Galactic potential models yield consistent values of $\dot{P}_{\rm b}^{\rm ext} \equiv \dot{P}_{\rm b}^{\rm Shk} + \dot{P}_{\rm b}^{\rm Gal}$.
Taking into account all terms, we use Eq.\,(\ref{eq:pbobs_eq}) to derive the intrinsic value of the variation of the orbital period, $\dot{P}_{\rm b}^{\rm intr}$. 
We use the masses derived from $\dot{\omega}$ and $\gamma$ in Eq.\,(\ref{eq:pbdot}) to obtain the orbital decay predicted by GR, $\dot{P}_{\rm b}^{\rm GR}=-462.0(5)\,{\rm fs\,s^{-1}}$. Thus the values of $\dot{P}_{\rm b}^{\rm intr}$ from the three Galactic electron-density models are clearly consistent with $\dot{P}_{\rm b}^{\rm GR}$. This result is discussed in more detail below.

\begin{figure}[htbp]
    \centering    \includegraphics[width=0.49\textwidth]{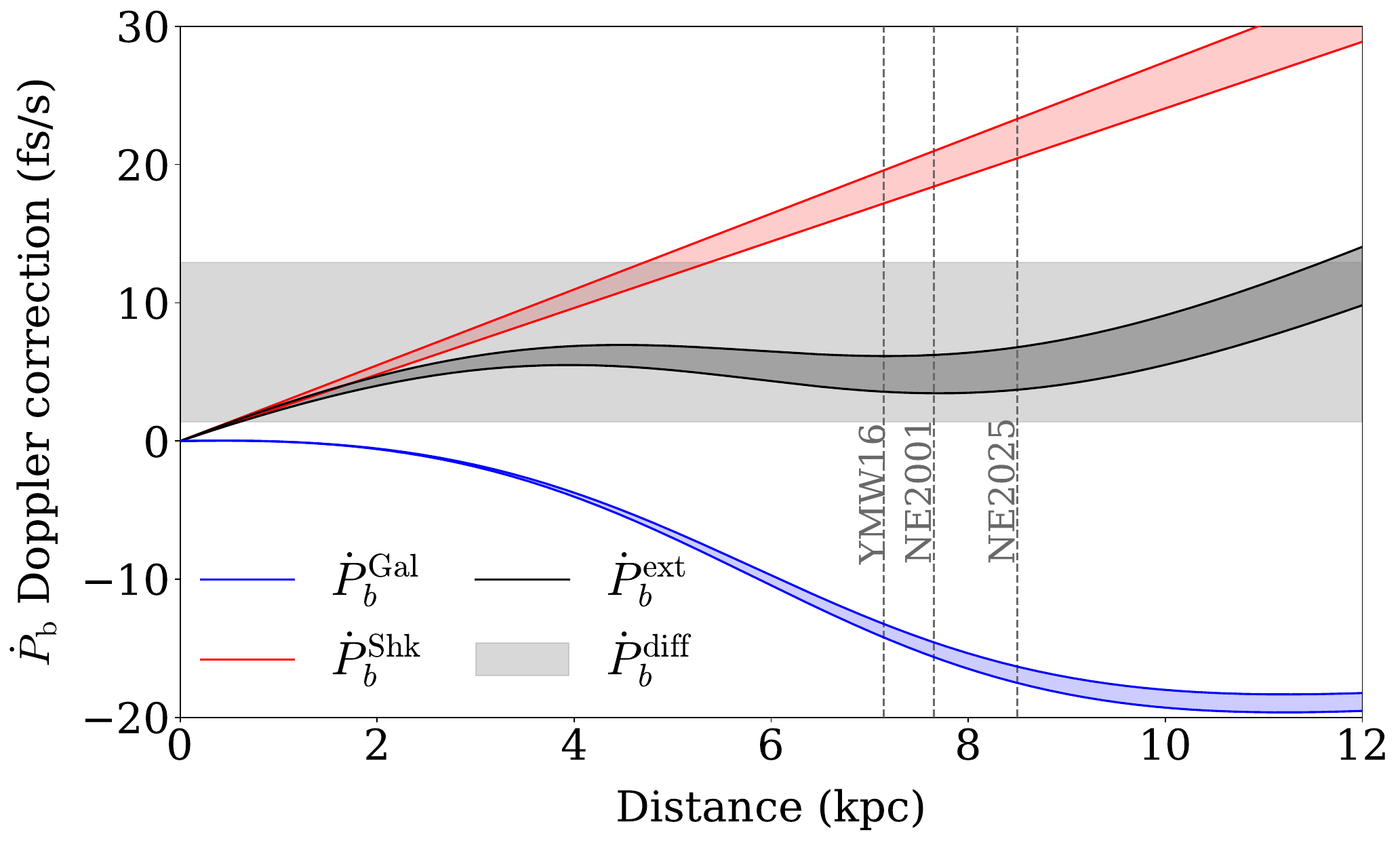}
    \caption{$\dot{P}_{\rm b}$ contributions from differential Galactic acceleration (\texttt{MWPotential2014} model) and the Shklovskii effect as a function of the distance to the pulsar. 
    The red curves denote the $\pm 1$-$\sigma$ limits of $\dot{P}_{\rm b}^{\rm Shk}$; the blue curves represent the same for $\dot{P}_{\rm b}^{\rm Gal}$; and the black curves indicate their sum, the total non-intrinsic contribution, $\dot{P}_{\rm b}^{\rm ext}$.
    The shaded regions between them indicate the 1-$\sigma$ uncertainties. The three vertical gray dashed lines mark the distances inferred from the YMW16, NE2001, and NE2025 models.
    The light gray band represents $\dot{P}_{\rm b}^{\rm diff}\equiv\dot{P}_{\rm b}^{\rm obs}-\dot{P}_{\rm b}^{\rm GR}$. \label{fig:pshk_pgal_gp}}
\end{figure}

\begin{figure*}[htbp]
    \centering
    \includegraphics[width=0.8\textwidth]{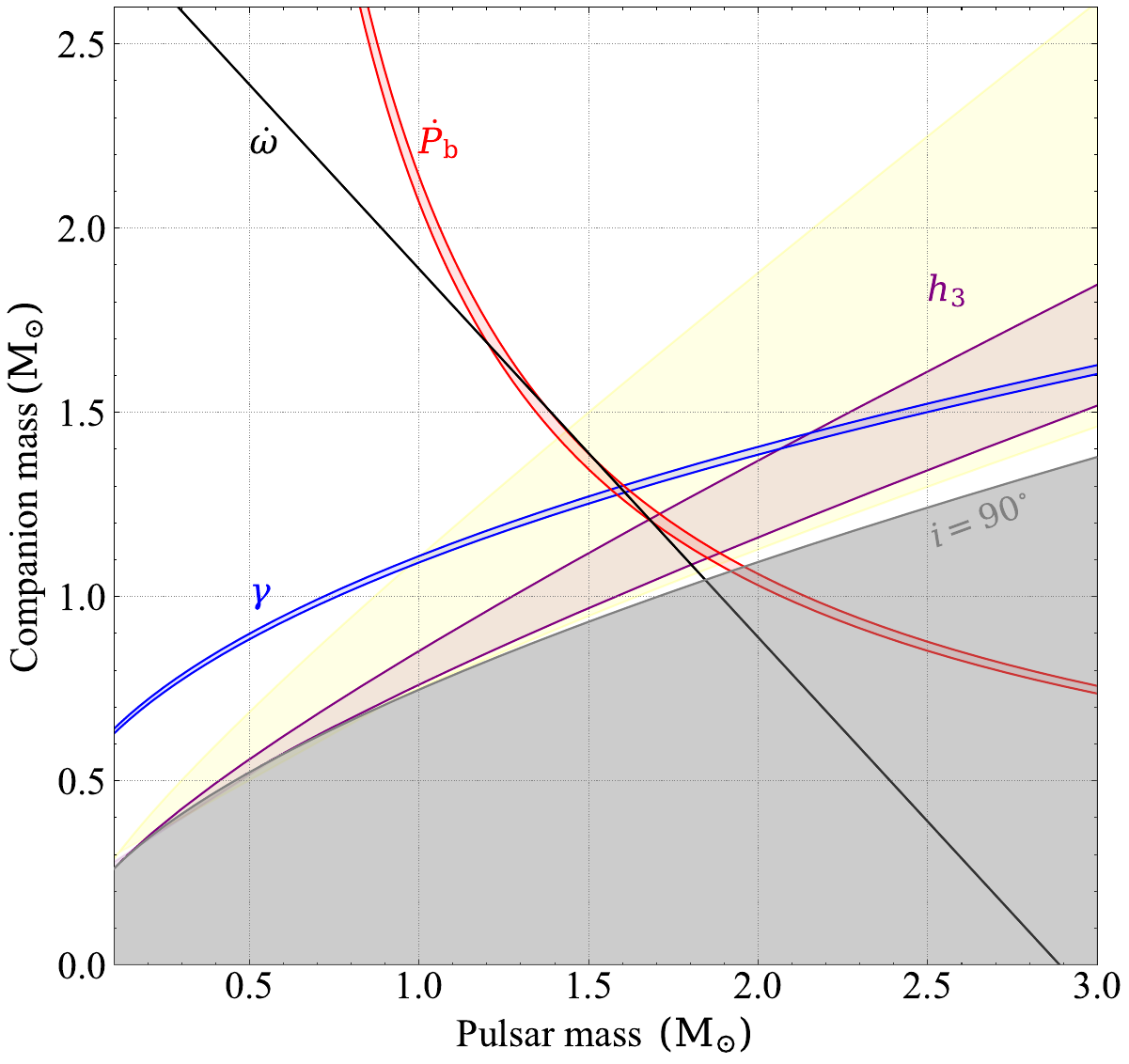}
    \caption{Mass-mass diagram for PSR J1913+1102. The measurements of the four PK parameters, $\dot{\omega}$, $\gamma$, $h_{3}$ and $\dot{P}_{\rm b}$, are displayed in different colors. The shaded regions represent the 1-$\sigma$ uncertainties of PK parameters, except for the yellow region, which indicates the 2-$\sigma$ uncertainty of $h_{3}$.
    $\dot{P}_{\rm b}$ has been corrected for the Shklovskii effect and the differential Galactic acceleration between the SSB and pulsar system. The gray area is excluded by the requirement that $\sin{i}\leq1$.\label{fig:mcmp_gp} }
\end{figure*}

In Fig.\,\ref{fig:pshk_pgal_gp}, the non-intrinsic contributions to the observed orbital period derivative are shown as functions of pulsar distance. The $\dot{P}_{\rm b}^{\rm Gal}$ term is calculated using the \texttt{MWPotential2014} Galactic potential model. The sum of these contributions ($\dot{P}_{\rm b}^{\rm ext}$) is represented by the dark gray region between solid black curves, denoting the 1-$\sigma$ constraints. The light gray region represents $\dot{P}_{\rm b}^{\rm diff}\equiv\dot{P}_{\rm b}^{\rm obs}-\dot{P}_{\rm b}^{\rm GR}$;
for the estimated distances we see that $\dot{P}_{\rm b}^{\rm diff}$ is compatible with $\dot{P}_{\rm b}^{\rm ext}$. 

One of the most important results in this paper is the finding that
the variation of $\dot{P}_{\rm b}^{\rm ext}$ across the pulsar distances inferred from the YMW16, NE2001 and NE2025 models is now relatively small. This is a result of our new measurement of the proper motion and the resulting improved estimate of the Shklovskii effect; its variation almost mirrors that of the Galactic acceleration in this region. Furthermore, the three Galactic electron-density models yield distance values that are consistent within the 1-$\sigma$ uncertainty. Assuming that they are correct, $\dot{P}_{\rm b}^{\rm ext}$ can be estimated with an uncertainty that is currently $\sim$3 times smaller than that of $\dot{P}_{\rm b}^{\rm obs}$, which implies that the uncertainty of $\dot{P}_{\rm b}^{\rm int}$ is still dominated by $\dot{P}_{\rm b}^{\rm obs}$.

We also correct the kinematic effects on $\dot{P}^{\rm obs}$ to obtain the intrinsic value. The formula is analogous to Eq.\,(\ref{eq:pbobs_eq}): $\left(\frac{\dot{P}}{P}\right)^{\rm intr} = \left(\frac{\dot{P}}{P}\right)^{\rm obs} - \left(\frac{\dot{P}}{P}\right)^{\rm Shk} - \left(\frac{\dot{P}}{P}\right)^{\rm Gal}$. We then use the intrinsic spin-down rate $\dot{P}^{\rm intr}$ to derive the characteristic age $\tau_{c}$, surface magnetic field $B_{S}$, and spin-down luminosity $\dot{E}$, which are also listed in Tab.\,\ref{tab:timing_gp}.

\section{Masses and gravity tests}\label{sec:mass_grav}
In this work, we obtain four PK parameters using the DDFWHE model. In any gravity theory, these parameters depend on the Keplerian parameters, the binary masses, and additional body-specific parameters, which depend on the specific gravity theories and generally, the equation of state (EoS) of dense matter. Under the assumption of GR, however, the PK parameters depend only, to leading order, on the Keplerian parameters and the component masses. Thus, with precisely measured Keplerian parameters, the binary masses can be determined using any two PK parameters, while the remaining two serve as independent consistency tests of gravity theories.

Figure\,\ref{fig:mcmp_gp} shows the GR-derived mass-mass diagram for PSR J1913+1102. 
Since the Shapiro parameter $\varsigma$ is fixed, we only show $h_{3}$ in the mass-mass diagram. 
In this figure, the PK parameters $\dot{\omega}$, $\dot{P}_{\rm b}$, and $\gamma$ overlap within their 1-$\sigma$ uncertainties, indicating consistency with GR at the 1-$\sigma$ level.
Compared with these three parameters, $h_{3}$ shows a larger deviation from the expected common intersection region.
Although $h_{3}$ does not overlap with the other three parameters within the 1-$\sigma$ uncertainty region, it overlaps with them within the 2-$\sigma$ uncertainty region, depicted by the yellow shaded area. This indicates that the difference from the GR prediction is not statistically significant.

\subsection{Masses of the two neutron stars}
In Sec.\ref{sec:omegadot}, we assumed GR to derive the total mass of this system, $m_{\rm tot}=2.88948(20)\,{\rm M_{\odot}}$, from the measured $\dot{\omega}$ using the DDFWHE model.
Combining $\dot{\omega}$ with the $\gamma$, the next most precisely measured PK parameter, yields the component masses $m_{\rm p}=1.599(8)\,\rm M_{\odot}$ and $m_{\rm c}=1.290(8)\,\rm M_{\odot}$.
We also applied the DDGR model, with the timing results summarized in the second column of Tab.\,\ref{tab:timing_gp}.
This model gives the total mass $m_{\rm tot}=2.88965(17)\,\rm M_{\odot}$, the pulsar mass $m_{\rm p}=1.599(8)\,\rm M_{\odot}$ and the companion mass $m_{\rm c}=1.291(8)\,\rm M_{\odot}$. 
The total mass and the individual component masses derived from the DDGR model are consistent, within 1 $\sigma$, with those obtained from the DDFWHE model. In the DDGR model, we derive $\sin i =0.811(5)$, corresponding to an inclination $i=54.2(5)^{\circ}$ (or $i=125.8(5)^{\circ}$). 

\begin{figure*}[htbp]
    \centering
    \includegraphics[width=0.75\textwidth]{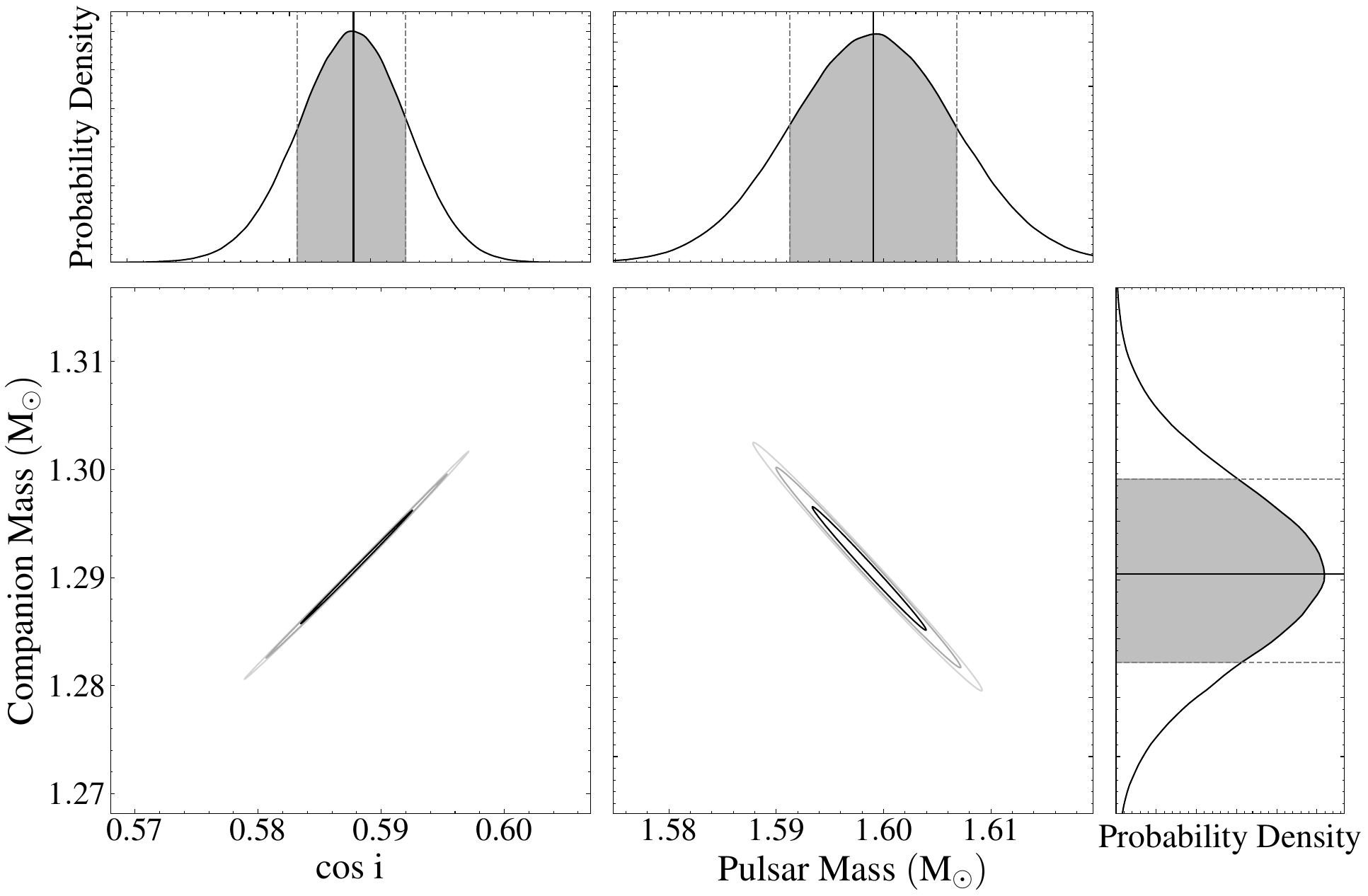}
    \caption{Maps of the fitted $\chi^{2}$ distributions and the corresponding probability density functions for the orbital inclination and the component masses in the DDGR model. The left panel shows the $\chi^{2}$ map as a function of $m_{\rm c}$ and $\cos i$. The right panel presents the corresponding $m_{\rm c}$-$m_{\rm p}$ distribution, translated into the $m_{\rm c}$-$m_{\rm p}$ plane via the mass function. The black, gray, and light gray contours correspond to the 68.27\%, 95.45\%, and 99.73\% confidence levels, which correspond to the 1, 2, and 3-$\sigma$ percentiles. The upper and rightmost panels show the marginalized one-dimensional probability distributions with marginalization. The gray regions represent the 1-$\sigma$ confidence region, and the black solid lines mark the median value, and the gray dashed lines are the $\pm1$-$\sigma$ values.\label{fig:mcmp_chisqure_gp}}
\end{figure*}

To refine the mass measurements, we also employed a Bayesian approach to compute the $\chi^{2}$ map for a grid of values of total mass $m_{\rm tot}$ and $\cos i$.
For each point in the grid, these values are kept fixed in a fit of a DDGR model to the timing data, the resulting $\chi^{2}$ values are then recorded. From these, we calculate the two-dimensional joint probability distributions, following a method similar to \citet{Splaver:2002ApJ}, and those are then projected into $m_{\rm c}$-$\cos i$ and $m_{\rm c}$-$m_{\rm p}$ spaces. We then calculate the marginalized probability distributions of $m_{\rm p}$, $m_{\rm c}$ and $\cos{i}$. We quote the posterior median as the central value and the 1-$\sigma$ confidence interval as the uncertainty: $m_{\rm p}=1.599(8)\,{\rm M_{\odot}}$, $m_{\rm c}=1.290(8)\,{\rm M_{\odot}}$, and $\cos{i}=0.588(7)$, corresponding to an inclination of $i=54.0(5)^{\circ}$ (or $i=126.0(5)^{\circ}$). Figure\,\ref{fig:mcmp_chisqure_gp} shows these probability distributions near $i=54.0(5)^{\circ}$. These values are consistent within 1 $\sigma$ with the results derived from the DDGR model.

\subsection{Testing GR with the orbital decay}
In Sec.\,\ref{sec:gamma}, we used $\dot{\omega}$ and $\gamma$ from the DDFWHE model to derive the component masses; using those masses we calculated the GR prediction for the orbital decay from GR emission, $\dot{P}_{\rm b}^{\rm GR}=-4.620(5)\times10^{-13}\,{\rm s\,s^{-1}}$. 
The intrinsic values of $\dot{P}_{\rm b}^{\rm intr}=-4.60(6)\times10^{-13}\,{\rm s\,s^{-1}}$, listed in Tab.\,\ref{tab:extralpbdot_dmgp}, are similar under the NE2001, YMW16 and NE20025 Galactic electron-density models. 
Hence, the ratio of $\dot{P}_{\rm b}^{\rm intr}$ to $\dot{P}_{\rm b}^{\rm GR}$ is given by
\begin{align}
    \frac{\dot{P}_{\rm b}^{\rm intr}}{\dot{P}_{\rm b}^{\rm GR}}=\frac{-4.60(6)\times10^{-13}}{-4.620(5)\times10^{-13}}=0.996(13)\,.
\end{align}
The $\dot{P}_{\rm b}^{\rm intr}$ is consistent with $\dot{P}_{\rm b}^{\rm GR}$ within the 1.3\% relative 1-$\sigma$ uncertainty. This is the fourth best measurement of the GR quadrupole formalism, after PSR~J0737$-$3039A/B, where the 1-$\sigma$ relative uncertainty is 0.0063\% \citep{Kramer:2021}, PSR~J1946+2052, where it is 0.09 \% \citep{Meng:2025} and PSR~B1913+16 (the ``Hulse-Taylor'' pulsar), where \cite{Weisberg:2016ApJ} quote a 1-$\sigma$ relative uncertainty of 0.16\% (but see \citealt{Deller:2018} for a more conservative error estimate). 

As we extend the timing baseline $T$, the uncertainty in $\dot{P}_{\rm b}^{\rm obs}$ will decrease as $T^{-5/2}$, as will the uncertainty in $\dot{P}_{\rm b}^{\rm intr}$.
This will lead to further improvements in this radiative test of GR.
Eventually, the precision will be limited by the uncertainty in $\dot{P}_{\rm b}^{\rm ext}$, as is already the case for PSR~B1913+16 \citep{Deller:2018}. However, we note that there is still some room to further improve the precision of $\dot{P}_{\rm b}^{\rm ext}$ through more accurate measurements of the proper motion. This and the improvement of $\dot{P}_{\rm b}^{\rm obs}$ will, in turn, enable a corresponding improvement in the estimate of $\dot{P}_{\rm b}^{\rm intr}$.

As we discuss in the following subsection, the mass asymmetry observed in PSR~J1913+1102 and the particular mass of the pulsar means that its measurement of the orbital decay plays an important role in testing scalar-tensor theories of gravity.

\subsection{Testing Damour-Esposito Far\`ese gravity}
GR has successfully passed all experimental tests to date, describing gravity as a long-range interaction mediated solely through a symmetric tensor field, the metric that describes the
spacetime geometry. 
Nevertheless, alternative theories of gravity have been studied for various reasons \citep[see e.g.][]{Clifton:2012PhR,Berti:2015,Will:2018book}. 
Many of these theories introduce additional scalar or vector degrees of freedom, among which the mono-scalar-tensor theory proposed by Damour and Esposito-Far$\grave{\rm e}$se \citep[DEF; ][]{Damour:1992CQGra,Damour:1993PhRvL,Damour:1996Prd} is particularly well studied.
In DEF gravity, the field equations can be expressed in the Einstein frame as,
\begin{align}
& R_{\mu \nu}^{*}=2 \partial_\mu \varphi \partial_\nu \varphi + \frac{8 \pi G_*}{c^4}\left(T_{\mu \nu}^{*}-\frac{1}{2} T^{*} g_{\mu \nu}^{*}\right)\,, \\
& g_{*}^{\mu \nu} \nabla_\mu^{*} \nabla_\nu^{*} \varphi=-\frac{4 \pi G_{*}}{c^4} \,\alpha(\varphi)\,T^{*}\,, \label{eq:FEphi}
\end{align}
where $g_{*}^{\mu \nu}$ is the metric tensor, $R_{\mu \nu}^{*}$ is the Ricci tensor, $T_{\mu \nu}^{*}$ is the stress-energy tensor and $T^{*}$ is the trace of $T_{\mu \nu}^{*}$, where ``$*$'' indicates quantities in the Einstein frame. 
$G_{*}$ is the bare gravitational, and the coupling strength $\alpha(\varphi)$ in Eq.~(\ref{eq:FEphi}) is given by
\begin{align}
   \alpha(\varphi) = \alpha_{0} + \beta_{0}(\varphi-\varphi_{0})\,,
\end{align}
where $\varphi_{0}$ denotes the asymptotic value of $\varphi$ at infinity, where one has $\alpha(\varphi_0) = \alpha_{0}$ and $\beta(\varphi_{0}) = \partial\alpha(\varphi) / \partial\varphi|_{\varphi_{0}} = \beta_{0}$. The Newtonian gravitational constant $G_\mathrm{N}$, as measured in a Cavendish experiment, is given by $G_{\rm N} = G_{*}(1+\alpha_{0}^{2})$.

In the weak-field limit, $\alpha_{0}$ and $\beta_{0}$ fully characterize the coupling of the scalar field $\varphi$ to matter. However, for strongly self-gravitating bodies such as pulsars, the weak-field approximation is not applicable. In this case, an effective scalar coupling strength $\alpha_{\rm A}$ must be used in place of $\alpha_{0}$, and its derivative $\beta_{\rm A}$ replaces $\beta_{0}$, where subscript ``A'' labels the body in the system. The definitions of $\alpha_{\rm A}$ and $\beta_{\rm A}$ are,
\begin{align}
\alpha_{\rm A} &= \frac{\partial\ln{m_{\rm A}}}{\partial\varphi_{a}}\,,\label{eq:alpha_a}\\
\beta_{\rm A} &= \frac{\partial{\alpha_{\rm A}}}{\partial\varphi_{a}}\,,
\end{align}
where $\varphi_{a}$ is the asymptotic scalar field which is felt by the considered body A. 
In Eq.\,(\ref{eq:alpha_a}), the value of $\alpha_{\rm A}$ depends on the mass of body A. For compact objects in DEF theory, there exists a critical baryonic mass $\bar{m}_{\rm cr}$ that marks the boundary between two regimes. For a total baryon mass $\bar{m}_{\rm A}<\bar{m}_{\rm cr}$, the effective scalar coupling strength $\alpha_{\rm A}$ remains small and is proportional to the weak-field limit. However, when $\bar{m}_{\rm A}>\bar{m}_{\rm cr}$, $\left|\alpha_{\rm A}\right|$ can reach order unity and becomes nearly independent of $\alpha_{0}$ \citep{Damour:1996Prd,Yagi:2021PhRvD}. 
This non-perturbative phenomenon, known as \textit{spontaneous scalarisation}, allows the theory to reduce to GR in the weak-field regime but to deviate significantly in strong gravitational fields \citep{Damour:1993PhRvL,Damour:1996Prd}. Similar behaviors also arise in other gravity theories \citep{Doneva:2024}.
Based on this characteristic, weak-field tests cannot properly constrain the effective scalar coupling parameter, making it necessary to probe gravity in the strong-field regime to test the DEF gravity and other scalar-tensor theories.
With at least one strongly self-gravitating NS, a binary pulsar system serves as an ideal laboratory to measure or constrain the non-GR parameter $\alpha_{\rm A}$ in DEF gravity.

In DEF gravity and other scalar-tensor frameworks, the energy loss of a system is carried not only by spin-2 GWs but also by spin-0 scalar waves. 
Among the scalar field contributions, dipolar radiation---arising at the 1.5 post-Newtonian (PN) order---can play a particularly significant role in GW emission \citep{Mirshekari:2013PhRvD}. 
The dipolar contribution is expressed as
\begin{align}
    \dot{P}_{\rm{b}}^{\text {dipole }}=-\frac{4 \pi^2 G_{*}}{c^3 P_{\rm{b}}} \frac{m_{\rm{p}} m_{\rm{c}}}{m_{\rm{p}}+m_{\rm{c}}} \frac{1+e^2 / 2}{\left(1-e^2\right)^{5 / 2}}\left(\alpha_{\rm{p}}-\alpha_{\rm{c}}\right)^2,
\end{align}
where $\alpha_{\rm{p}}$ and $\alpha_{\rm{c}}$ are the effective scalar couplings of the pulsar and the companion respectively \citep{Damour:1996Prd}. With precise measurements of $\dot{P}_{\rm b}$ and the component masses, one can either determine or place a stringent constraint on the effective scalar coupling parameter. 
For a pulsar-white dwarf (WD) system, since the WD is a weakly self-gravitating body compared to the NS, its effective scalar coupling parameter satisfies $\alpha_{\rm A} \simeq \alpha_{0}$. Consequently, the constraint on $\left|\alpha_{\rm p}-\alpha_{\rm c}\right|$ can be replaced by a constraint on $\left|\alpha_{\rm p}-\alpha_{0}\right|$.

In contrast, for a PSR-NS system, both $\alpha_{\rm p}$ and $\alpha_{\rm c}$ depend significantly on the EoS; furthermore,  if the masses of the two NSs are similar, then $\left|\alpha_{\rm p}-\alpha_{\rm c}\right|$ will be quite small irrespective of the EoSs. For this reason, PSR-NS systems are generally not ideal for this sort of tests. 

However, the large mass asymmetry of PSR~J1913+1102 implies that $\left|\alpha_{\rm p}-\alpha_{\rm c}\right|$ could be unusually large for a PSR-NS system, making this an unusually sensitive system for deviations from GR. Furthermore, the mass of the pulsar ($\sim 1.6 \, \rm M_{\odot}$) places it in a region of the NS mass spectrum where no strong constraints on $\alpha_{\rm p}$ existed beforehand.

Based on the framework developed in \citet{Shao:2017PRX}, \citet{Zhao:2022CQGra} used the latest timing results from seven well-timed binary pulsar systems, including PSR J1913+1102, to constrain the effective scalar coupling $\alpha_{\rm A}$.
They focused on limiting the spontaneous-scalarisation window, corresponding to $\beta_{0} \in [-4.8, 4.0]$.
Through a Bayesian analysis, they obtained the constraint $\left|\alpha_{\rm A}\right| \lesssim 6\times10^{-3}$, thereby effectively closing the spontaneous-scalarisation window for DEF gravity.
As shown in Fig.\,2 of \citet{Zhao:2022CQGra}, PSR J1913+1102 plays an important role in constraining the effective scalar coupling parameter around this mass range.

In our work, we improved the precision of the measurements of $m_{\rm p}$, $m_{\rm c}$, and $\dot{P}_{\rm b}$ for PSR J1913+1102; this should result in improved constraints on DEF gravity. To verify this, we adopted the method of \citet{Zhao:2022CQGra} to further constrain $\alpha_{\rm A}$ within the spontaneous-scalarisation window.

\begin{figure}
    \centering
    \includegraphics[width=0.49\textwidth]{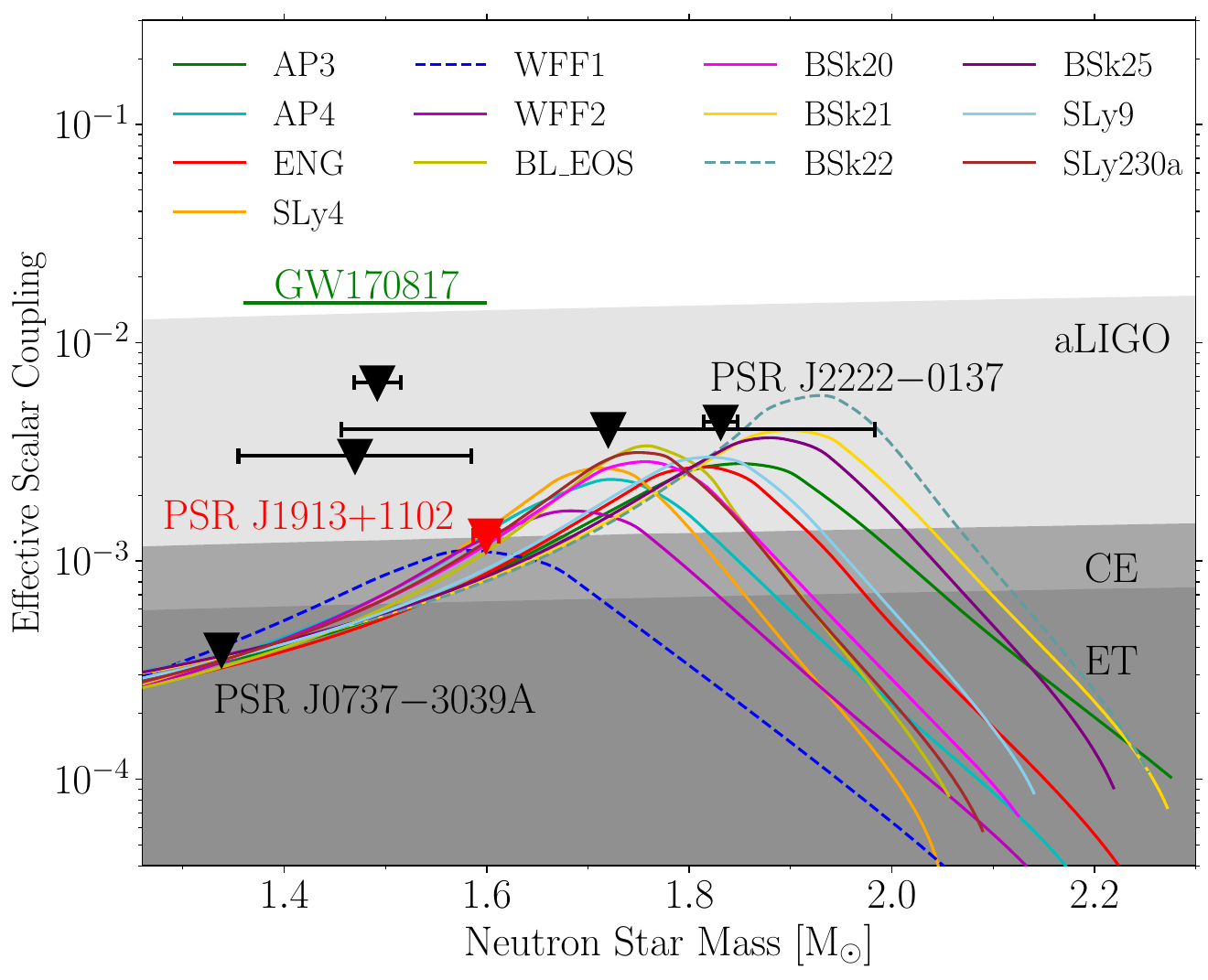}
    \caption{Updated constraints on the effective scalar coupling parameter $\alpha_{\rm A}$ in the spontaneous scalarisation window ($\beta_{0}\in\left[-4.8,\,4.0\right]$), as a function of NS's mass, derived from a set of binary pulsar systems and for different NS's EoSs. The inverted triangles represent the $90\%$ CL upper limits on the effective scalar coupling parameter $\alpha_{\rm A}$ from a selection of binary pulsars. The updated constraints from PSR J1913+1102, which incorporate the new timing results of this pulsar, are indicated by the red triangle.
    The coloured lines represent the 90\% limits obtained from the analysis of the ensemble of pulsars obtained assuming the EoSs listed at the top. The figure also highlights PSRs J2222$-$0137 and J0737$-$3039A, which together with PSR J1913+1102 currently play the most important role in constraining the DEF spontaneous-scalarisation window. We also show the constraint on $\alpha_{\rm A}$ from GW170817.
    The shaded gray regions indicate parameter space below the sensitivity of advanced LIGO (aLIGO) and the projected sensitivities of Cosmic Explorer (CE) and Einstein Telescope (ET), obtained via Fisher-matrix estimates for a $1.25\,{\rm M_{\odot}}$ companion NS merger at 200 Mpc ($90\%$ CL upper limits).} \label{fig:scalarwindow}
\end{figure}

Figure\,\ref{fig:scalarwindow} shows the updated constraints on $\alpha_{\rm A}$ in the spontaneous-scalarisation window, evaluated for different EoSs, based on six binary pulsar systems (excluding PSR J0348+0432\footnote{Recent timing measurements of PSR J0348+0432 \citep{Saffer:2025ApJ,JP:2025} yield a mismatch between the $\dot{P}_{\rm b}^{\rm obs}$ and the GR prediction for the masses measured by \cite{Antoniadis:2013}; the reason for this is not yet fully understood. For this reason, we have excluded it from our sample.}) selected by \citet{Zhao:2022CQGra}.
As illustrated in Fig.\,\ref{fig:scalarwindow}, incorporating the new timing results for PSR J1913+1102 (marked by a red triangle) leads to a substantial improvement in the constraints on $\alpha_{\rm A}$ around $1.6\,{\rm M_{\odot}}$, relative to the right-hand panel of Fig.\,2 in \citet{Zhao:2022CQGra}. These new timing results therefore further constrain the magnitude of the DEF spontaneous-scalarisation phenomenon near $1.6 \, \rm M_{\odot}$.
In addition, Fig.\,\ref{fig:scalarwindow} marks PSRs J0737$-$3039A \citep{Kramer:2021} and J2222$-$0137 \citep{2021A&A:Guo}, which provide the strongest constraints at the low-mass and high-mass ends of the window, respectively.
Taken together, PSRs J0737$-$3039A, J2222$-$0137, and J1913+1102 currently play the most important roles in defining the allowed region of the DEF spontaneous-scalarisation window.
Building on this improvements of PSR J1913+1102, continued pulsar timing observations of this system, combined with the scaling $\dot{P}_{\rm b}^{\rm obs} \propto T^{-5/2}$ \citep{Damour:1992PrD}, will further tighten the constraints on $\alpha_{\rm A}$, and thus provide a more stringent closure of the DEF spontaneous-scalarisation window in the near future.

\section{Evolution of this system}\label{sec:evo}

\subsection{Motivation for studying DNS formation}

Apart from studies of gravity theories, DNSs have become important for several other reasons.
The first is how massive binary stars evolve and form DNSs. As mentioned above, of the 570 binary pulsar systems currently known, only 21 (3.7\%) are confirmed DNS systems. This naturally raises the question of why DNS systems are so rare. To address this question, we need to trace the evolutionary history of DNS systems back to their formation.
This is important because a good understanding of their abundance (and orbital characteristics) yields estimates of DNS merger rates \citep[see][and references therein]{Miquel_2023}, which can be compared to what is observed by ground-based detectors like LIGO/Virgo/Kagra \citep{LVK:2025}. To date, the latter have seen two DNS merger events, GW170817 in the second observing run \citep[O2;][]{TheLIGOScientific:2017qsa} and GW190425 in the third run \citep[O3; ][]{Abbott:2020uma}. Looking ahead, space-based detectors --- including LISA \citep{LISA23}, TianQin \citep{lbb+25}, and Taiji \citep{Luo:2021PTEP} --- are expected to detect low-frequency GWs from Galactic DNS systems up to $\sim1\,{\rm Myr}$ before merger, complementing pulsar observations in constraining DNS formation, NS masses, and merger rates.

Theoretical investigations to date have established a general framework for understanding how DNS systems form, and Fig.\,1 of \citet{Tauris:2017} gives an illustration of the formation of a DNS system. 
Starting from an OB-star binary, the system undergoes multiple stages of mass transfer, one or more common-envelope (CE) episodes, and two SN explosions, and ultimately survives as a DNS system.
However, several uncertainties remain in the formation of DNS systems, including whether {\em all} systems undergo an additional phase of mass transfer --- Case~BB Roche-lobe overflow \citep{Savonije:1976A&A,Greve:1977Ap&SS,Delgado:1981A&A,Tauris:2015MNRAS} --- before the second SN, which allows for extreme stripping of the helium star prior to collapse \citep{Tauris:2013ApJ,Tauris:2015MNRAS,Suwa:2015MNRAS,Moriya:2017MNRAS,jtcf21}. 
Additional uncertainties concern how momentum kicks (magnitude and direction) are imparted to newborn NSs, and whether the second SN typically imparts kicks that are, on average, substantially smaller than those of the first SN. Estimating these kick magnitudes, and how they relate to the masses of the NSs and their progenitors (see discussion in \citealt{Tauris:2017}) is particularly important for the study of SN physics.
In this context, observational constraints are crucial for reducing these uncertainties and testing theoretical predictions.
Radio-detectable DNS systems play an essential role in tracing back their formation. Their spin periods, spin-period derivatives, orbital periods, eccentricities, NS masses, locations, and systemic velocities provide crucial diagnostics for exploring kick distributions and the masses of the NSs and of their pre-SN progenitors.

\subsection{Kinematic traceback of PSR J1913+1102}

In their global study of the properties of NSs, \citet{Tauris:2017} used timing results from 15 DNS systems, including PSR J1913+1102, to conduct a comprehensive analysis of their characteristics at formation, including numerous MC simulations of the second SN to infer pre-SN stellar properties and probe the explosion physics.
Compared with the data used by \citet{Tauris:2017}, our improved NS mass and proper motion measurements allow us to repeat their MC framework and update the constraints on the pre-SN properties of PSR J1913+1102.

The updated proper motion measured in this work, combined with the NE2001 distance estimate, yields a transverse velocity $v_{\rm T}^{\rm SSB}=279(56)\, \rm km\,s^{-1}$ in the SSB frame,
following the method and treatment of uncertainties outlined in \citet{Tauris:2017}. To infer the full 3D velocity, we impose a random vector orientation and thus determine the radial component $v_{\rm R}^{\rm SSB}$. With both components in hand, we apply the \texttt{MWPotential2014} Galactic potential to convert the SSB-frame velocity vector to LCF frame, yielding a systemic velocity within the 1-$\sigma$ interval $v_{\rm sys} =65$--$309\,{\rm km\,s^{-1}}$.

\begin{figure}[htbp]
    \centering
    \includegraphics[width=0.5\textwidth]{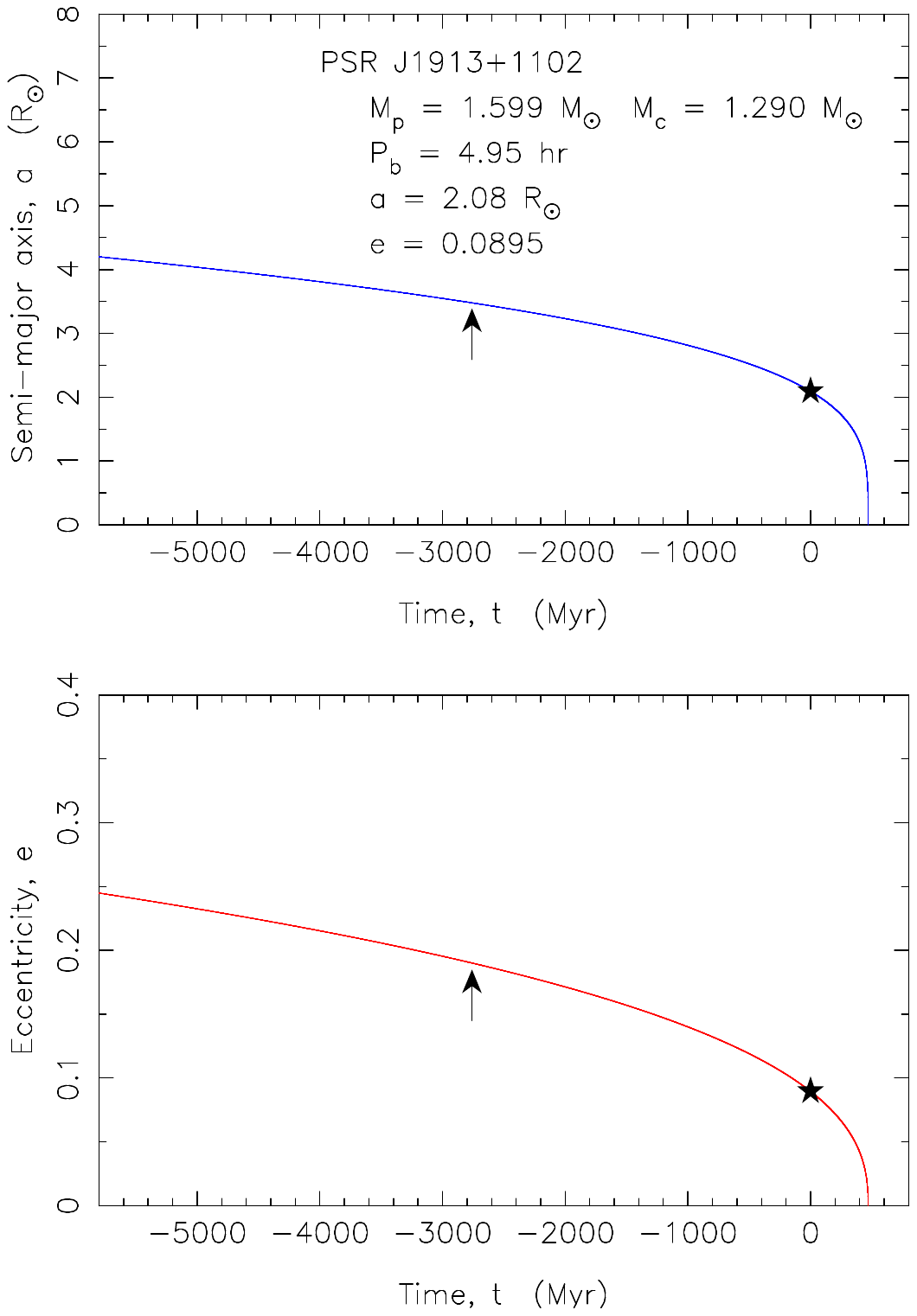}
    \caption{Past ($t<0$) and future ($t>0$) evolution of the semi-major axis (top panel) and eccentricity (bottom panel) of PSR J1913+1102. The black star marks the current system at $t=0$, and the annotations in the top panel display the parameter values at the current epoch.
    Assuming a braking index of $n=3$ and post-SN spin period of $P_{\rm SN}\rightarrow 0$, the maximum age of this system is $\sim 2730\;$Myr, which is indicated by the black solid arrow. The parameters are evolved back to the post-SN parameters of PSR J1913+1102.\label{fig:orbit_timevari}}
\end{figure}

PSR J1913+1102 is a relativistic DNS system whose orbit undergoes significant shrinkage due to GW damping ($\dot{a}\simeq 0.35\,{\rm m\,yr}^{-1}$). 
Therefore, it is necessary to map the observed system back to its post-SN state.
Using the approach outlined in Sec.\,7 of \citet{Tauris:2017} and adopting the lowest-order secular evolution of a two-point-mass binary \citep{Peters:1964}, we trace the system back to its post-SN parameters, assuming a braking index of $n=3$ and letting the post-SN spin period $P_{\rm SN}\rightarrow 0$ to determine an upper limit on its age.
Fig.\,\ref{fig:orbit_timevari} shows the past and the future evolution of PSR J1913+1102 in terms of its semi-major axis and orbital eccentricity. 
The black star marks the current epoch of PSR J1913+1102 at $t=0$, where $a=2.09\,R_{\odot}$ and $e=0.0895$; the black solid arrow indicates the trace-back to the post-SN orbital parameters.

\subsection{Supernova kick constraints from Monte Carlo simulations}
\begin{figure*}[htbp]
    \centering
    \includegraphics[width=0.93\textwidth]{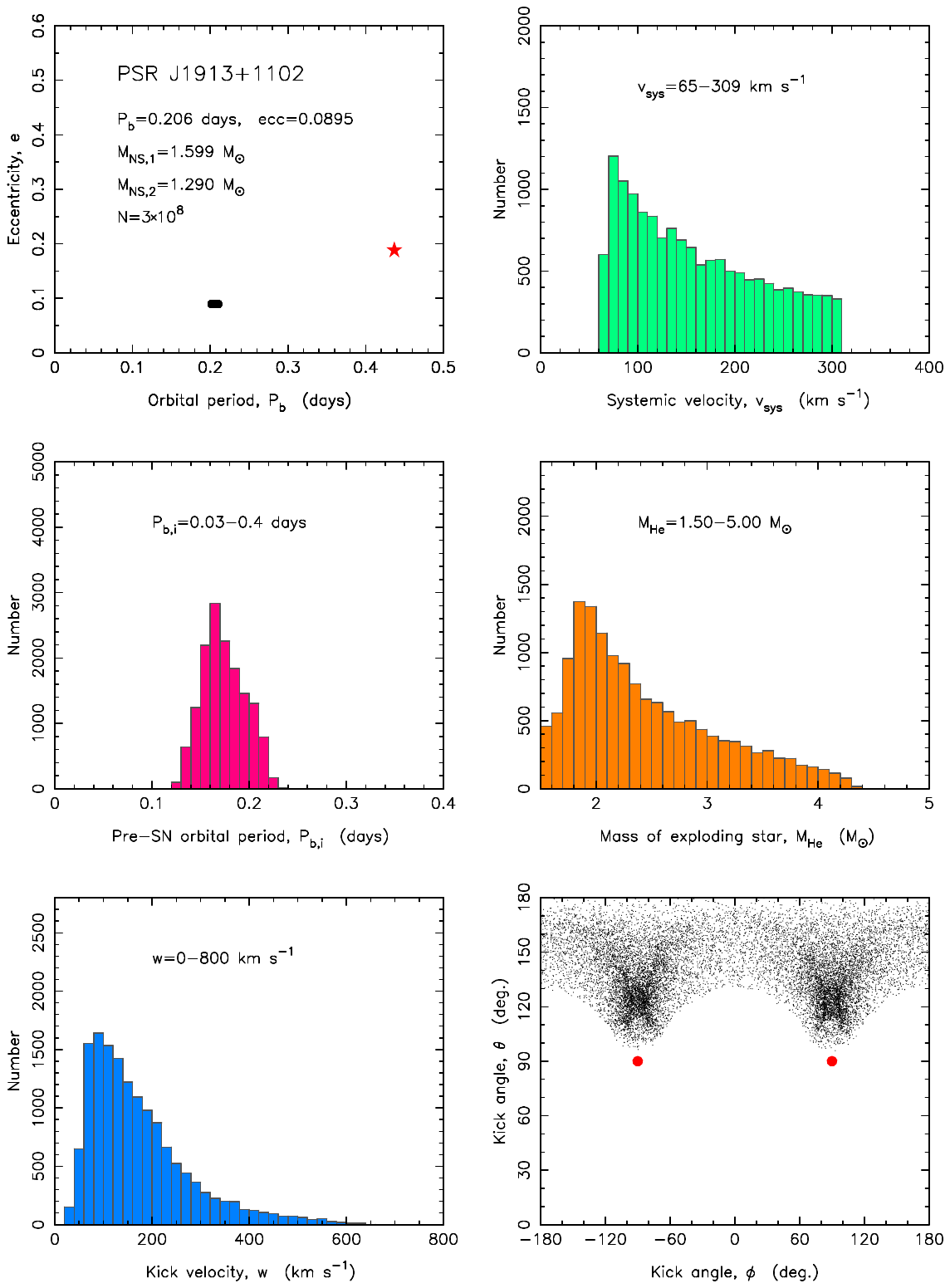}
    \caption{Formation constraints of PSR J1913+1102 based on $3\times 10^8$ simulations of the second SN. Here, $v_{\rm sys}$ is the 3D systemic velocity with respect to the LCF, $P_{\rm b,\,i}$ is the pre-SN orbital period, ${\rm M}_{\rm He,\,f}$ is the final helium-star mass of the exploding progenitor, $w$ is the kick magnitude imparted to the newborn NS, and $\theta$ and $\phi$ are two angles defining the direction of the kick velocity. In the upper-left panel, the red star shows the initial location of this system if it formed 2730 Myr ago. In the lower-right panel, the red circles indicate kick directions parallel ($\theta=90^{\circ},\,\phi=90^{\circ}$) and anti-parallel ($\theta=90^{\circ},\,\phi=-90^{\circ}$) to the pre-SN orbital angular momentum vector. \label{fig:sn_limit}}
\end{figure*}

With these parameters of PSR J1913+1102, we apply the MC framework of \citet{Tauris:2017} to simulate SN explosions and compute the resulting kinematic properties of the system that survived it.
In the MC simulation, we explore a five-dimensional parameter space consisting of the final helium-star mass of the exploding progenitor (${\rm M}_{\rm He,\,f}$), the pre-SN orbital period ($P_{\rm b,\,i}$), the kick magnitude imparted to the newborn NS ($w$), and the two angles defining the kick direction ($\theta$, $\phi$) \footnote{Further details of the simulations can be found in \citet{Tauris:2017}.}.
Fig.\,\ref{fig:sn_limit} shows the results from the MC simulation. The upper-left panel displays the {initial post-SN values (red star) and a black square with accepted post-SN solutions from the MC simulations within a $3\%$ accepted error margin of the current values of $P_{\rm b}$ and $e$}.
The upper-right panel shows the posterior distribution of the 3D systemic velocities with respect to the LCF. 
The middle-left panel shows the distribution of pre-SN orbital period, $ 0.11\,{\rm days} \lesssim P_{\rm b,\,i}\lesssim 0.23\,{\rm days}$. The middle-right panel presents the distribution of the mass of the exploding star, with ${\rm M}_{\rm He,\,f}\lesssim 4.3\,{\rm M_{\odot}}$. A peak appears at ${\rm M}_{\rm He,\,f}<2\,{\rm M_{\odot}}$, suggesting that lower masses are statistically favored; however, the extended tail implies that values as high as ${\rm M}_{\rm He,\,f}\sim4\,{\rm M_{\odot}}$ cannot be excluded.
The lower-left panel shows the kick velocities for all solutions, spanning a wide range, $25\,{\rm km\,s^{-1}}\lesssim w \lesssim 600\,{\rm km\,s^{-1}}$, exhibiting a peak slightly below $100\,{\rm km\,s^{-1}}$.
The lower-right panel shows the two kick angles, $\theta$ and $\phi$. The red circles mark the directions parallel ($\theta=90^{\circ},\,\phi=90^{\circ}$) and anti-parallel ($\theta=90^{\circ},\,\phi=-90^{\circ}$) to the pre-SN orbital angular momentum vector. The kick-angle distribution shows a significant preference for backward directions ($\theta>90^{\circ}$), because such kicks increase the probability that the system remains bound.
Compared with Fig.\,35 of \citet{Tauris:2017}, the improved mass and proper-motion measurements lead to tighter constraints on the five parameters shown in Fig.\,\ref{fig:sn_limit}. However, although the new results further reduce the allowed parameter space, the current constraints --- both on ${\rm M}_{\rm He,\,f}$ and on $w$ --- are still insufficient to identify a unique SN solution and formation pathway for PSR~J1913+1102, or addressing general uncertainties in DNS formation in a broader context.

It is important to keep in mind that Fig.~\ref{fig:sn_limit} makes no
\emph{a priori} assumptions about the likelihood of the input parameter
distributions that affect the inferred post-SN parameters. The figure is
therefore intended to demonstrate purely the \emph{kinematically} allowed
solutions. Consequently, adopting more ``realistic'' exploding-star masses
(e.g. $M_{\rm He,f} \lesssim 2.0\;{\rm M_\odot}$) and natal kick magnitudes
($w \lesssim 100\;{\rm km\,s}^{-1}$) that produce a typical
$1.29\, \rm M_\odot$ NS \citep[][and references therein]{Tauris:2017} would
likely alter the relative likelihood of the inferred pre-SN parameters.

Nevertheless, until tighter constraints can be placed on the kinematics of
this system, no secure conclusions can be drawn regarding the origin of
PSR~J1913+1102. It should also be noted that Fig.~\ref{fig:sn_limit} shows
the outcome of simulations reproducing PSR~J1913+1102 as it is observed
\emph{today}. In principle, immediately after the second SN, the system may
have had orbital parameters corresponding to those indicated by the red star
in the upper-left panel, if it formed approximately $2730\;{\rm Myr}$ ago.
Even so, for this system it remains limited what can be learned by invoking
a wider and more eccentric post-SN orbit to account for subsequent GW
damping.

External geometric constraints from DNS systems can aid in constraining their formation, such as the misalignment angle --- the angle between the pulsar's spin axis and the orbital angular momentum vector --- that offers valuable quantitative constraints on the SN kick, 
as demonstrated by systems such as PSRs B1913+16, B1534+12, and J0737–3039A/B \citep{Tauris:2017}\footnote{An earlier analysis of the kinematics of PSR~B1913+16 utilizing the information of spin-orbit misalignment in this system was made by \cite{Wex:2000}. Note, however, that the values of distance and especially proper motion then available were substantially different from those later determined by \cite{Weisberg:2016ApJ}; this appears to be the main reason for the different conclusions on the dynamics of this system.}.
In principle, the evolution of the average pulse profile and linear polarization due to geodetic precession can be used to measure or constrain this angle. For PSR J1913+1102, however, severe pulse broadening in the L band due to interstellar scattering, together with the absence of detectable linear polarization, prevents a reliable determination of the misalignment angle. Consequently, this system cannot benefit from this geometric constraint to refine its binary-formation history.

\section{Conclusion and Prospects}\label{sec:con_pro}
\renewcommand{\arraystretch}{1.34}
\begin{table*}
\centering
\caption{Improvement of the precision of the PK parameters for PSR J1913+1102 relative to the parameters in this work. The simulations are divided into four time spans: 2012-2030 (pre-SKA AA*), 2012-2033 (including 3 years of SKA AA*), 2012-2035 (including 5 years of SKA AA*), and 2012-2040 (including 10 years of SKA AA*). The estimates of SKA AA* observations are at S-band (1650-3050\,MHz).
The last row gives the improvement of the precision of the $\dot{P}_{\rm b}^{\rm intr}$.
\label{tab:simulation_ska}}
\begin{tabular}{ccccc}
\hline\hline
~ & Pre-SKA AA* (2012-2030)  & Pre-SKA AA*  +  3 yr  & Pre-SKA AA* + 5 yr & Pre-SKA AA* + 10 yr  \\
\hline
$\dot{\omega}$ & 2.51 & 4.40 & 5.64 & 8.98 \\
$\gamma$ & 1.66 & 1.96 & 1.98 & 2.35 \\
$h_{3}$ & 1.79 & 2.93 & 3.54 & 4.80 \\
$\dot{P}_{\rm b}^{\rm obs}$&2.92 & 5.56 & 7.48 & 15.15 \\
$\dot{P}_{\rm b}^{\rm intr}$& 2.60 & 4.25 & 4.95 & 6.14 \\
\hline
\end{tabular}
\end{table*}

In this work, we present a comprehensive timing analysis of PSR J1913+1102 based on 13 years of observations from Arecibo and FAST.
Due to significant DM variations, we model them using a Gaussian process and verify the results with a 14-order polynomial fit. For each DM fitting method, we perform pulsar timing with TEMPO using the DDFWHE and DDGR binary models. No statistically significant difference was found between the two methods.

From the DDFWHE model we obtain measurements of four PK parameters---$\dot{\omega}$, $\gamma$, $\dot{P}_{\rm b}$, and $h_{3}$---using a fixed $\varsigma$. For $\dot{\omega}$, we achieve twice the precision reported by \citet{Ferdman:2020Natur}, which in turn results in a threefold improvement in the determination of the total binary mass $m_{\rm tot}$. A 3.7-fold improvement in the precision of $\gamma$ then leads to a 3.7-fold refinement in the determination of the component masses $m_{\rm p}$ and $m_{\rm c}$. Owing to the scaling of the uncertainty of $\dot{P}_{\rm b}^{\rm obs} (\propto T^{-5/2})$ and to the improved DM variation modeling, the incorporation of three additional years of FAST timing data has reduced the uncertainty in the observed orbital decay $\dot{P}_{\rm b}^{\rm obs}$ by a factor of five.
We have also achived a 3.3- and 3.7-fold improvement in the measurement of the two components of the proper motion.

The precisely determined component masses allow a precise prediction of the orbital decay due to the emission of GWs, $\dot{P}_{\rm b}^{\rm GR}$.
The improved proper motion measurement enables a more accurate estimate of $\dot{P}_{\rm b}^{\rm ext}$. By subtracting this contribution from the observed value, $\dot{P}_{\rm b}^{\rm obs}$, we obtain a measurement of the intrinsic orbital decay with a relative uncertainty of 1.3\%. 
Its comparison with $\dot{P}_{\rm b}^{\rm GR}$ represents a test of the GR quadrupole formula; the agreement between these two values means that GR has passed this test. More importantly, the asymmetric nature of this DNS system results in stringent limits on dipolar GW emission and thus on the scalar coupling parameter $\alpha_{\rm A}$ for $1.6\,{\rm M_{\odot}}$ NSs in the DEF gravity, highlighting the unique importance of PSR J1913+1102 for testing scalar-tensor gravity.

The improved proper motion measurement has another important consequence: with our measured magnitude of the proper motion, the value of $\dot{P}_{\rm b}^{\rm ext}$ remains nearly constant over a wide range of assumed distances (see Fig.\,\ref{fig:pshk_pgal_gp}), which reduces the uncertainty of $\dot{P}_{\rm b}^{\rm ext}$ to a level that is three times smaller than the uncertainty of $\dot{P}_{\rm b}^{\rm obs}$. This implies that further improving $\dot{P}_{\rm b}^{\rm obs}$ by continued timing will improve $\dot{P}_{\rm b}^{\rm int}$ by at least a factor of 3 before the uncertainty of $\dot{P}_{\rm b}^{\rm ext}$ becomes the limiting factor. Moreover, we note that the improvement in the precision of the proper motion in the near future will also further decrease the uncertainty of $\dot{P}_{\rm b}^{\rm ext}$. Thus, continued observations of PSR J1913+1102 by FAST in the future will significantly enhance the capacity of this system to constrain scalar-tensor theories of gravity.

The improved proper motion also allows an estimate of the transverse peculiar velocity, which also happens to vary slowly with distance.
Within the 1-$\sigma$ distance range inferred from three Galactic electron-density models, we obtain $\Delta v_{l}\sim -40\,{\rm km\,s^{-1}}$ and $\Delta v_{b}\sim 10\,{\rm km\,s^{-1}}$.
This relatively small transverse peculiar velocity suggests that the second SN imparted a relatively small kick.
This and the improved mass measurements help us constrain the origin of this DNS system. Using the measured parameters, we trace the system back to the second SN and simulate the explosion.
Based on $3\times10^{8}$ MC simulations, we obtain tighter constraints on the final helium-star mass (${\rm M_{He,\,f}}$) and the magnitude ($w$) and direction ($\phi,\,\theta$) of the second SN kick velocity.
However, these improvements remain insufficient to uniquely determine a SN solution and formation pathway of PSR J1913+1102, or to resolve the broader uncertainties associated with DNS formation.

Since L-band observations of PSR J1913+1102 are strongly affected by interstellar scattering and DM variations, potential future S-band observations with FAST or SKA-Mid\footnote{\url{https://zenodo.org/records/16895574}} are expected to provide more precise measurements of the PK parameters than L-band observations over the same time span.
In \citet{Krishnan:2025}, assuming SKA AA* and SKA AA4\footnote{For SKA-Mid, SKA AA* is described as an intermediate stage comprising 144 Mid dishes, while SKA AA4 corresponds to the full design baseline with 197 Mid dishes; further details are available at \url{https://zenodo.org/records/16951020}.} become operational after 2028, 10\,yr of simulated ToAs (2028-2038) at L- and S-band, combined with simulated historical Arecibo and FAST datasets, show that S-band yields significantly improved timing precision, particularly for $\dot{P}_{\rm b}$, with uncertainties reduced by a factor of $\sim$2 compared to L-band.
However, given updated expectations that SKA AA* will not be operational before 2031 and that SKA AA4 may not be realized, we perform revised simulations following \citet{Krishnan:2025}. Assuming S-band observations (1650-3050\,MHz) with a ToA uncertainty of 12\,$\upmu$s, we simulate 3, 5, and 10\,yr of SKA AA* data starting from 2031, combined with simulated Arecibo and FAST data from 2012-2025 and continued FAST observations from 2025-2030. 
For each case, we derive the PK parameters and compare their uncertainties with those from the current $\sim$13\,yr dataset; 
the improvement of the precision are summarized in Tab.\,\ref{tab:simulation_ska}. Focusing on orbital decay, the precision of $\dot P_{\rm b}^{\rm obs}$ improves significantly with increasing S-band baseline, reaching a factor of $\sim$15 for 10\,yr of SKA AA* relative to the current $\sim$13\,yr dataset. However, after $\sim$3 yr of SKA AA* observations, the uncertainty in $\dot P_{\rm b}^{\rm intr}$ is already dominated by the uncertainty in the system distance. Consequently, despite the substantial improvement in $\dot P_{\rm b}^{\rm obs}$, the precision of $\dot P_{\rm b}^{\rm intr}$ will eventually be limited by the uncertainty of $\dot P_{\rm b}^{\rm ext}$.

To summarize, there is still substantial scope for improving the precision of parameters relevant to tests of gravity with this system, as well as those constraining its formation and evolutionary history.
Moreover, future observational capabilities will enable a more rapid refinement of its timing precision.
We therefore anticipate that PSR J1913+1102 will continue to demonstrate its unique value as a natural laboratory for testing gravity and also improve our understanding of DNS.


\section*{Acknowledgements}
We thank the anonymous referee for constructive comments that significantly improved the work.
We thank Heng Xu and Dejiang Zhou for valuable discussions and suggestions.
We thank Cijie Zhang and Yukai Zhou for deploying and maintaining the server used throughout this research.
This work was supported by the National Key R$\&$D Program of China No. 2022YFC2205203 and the CAS-MPG LEGACY project.
XM was supported by the National Natural Science Foundation of China (12203072).
WZ was supported by the National Science Foundation of China (12421003), the Chinese Academy of Sciences Project for Young Scientists in Basic Research (YSBR-063), and the Beijing Nova Program (20250484786).
PCCF, NW, MK and DJC gratefully acknowledge continuing support from the Max-Planck-Gesellschaft.
LS was supported by the National SKA Program of China (2020SKA0120300), the National Natural Science Foundation of China (12573042), and the Max Planck Partner Group Program funded by the Max Planck Society.
This work made use of the data from FAST (Five-hundred-meter Aperture Spherical radio Telescope). 
FAST is a Chinese national mega-science facility, operated by National Astronomical Observatories, Chinese Academy of Sciences.
While taking data for this work, the Arecibo Observatory was operated by SRI International under a cooperative agreement with the U.S. National Science Foundation (NSF; AST-1100968), and in alliance with Ana G. M\'endez-Universidad Metropolitana, and the Universities Space Research Association. This work was partly based on observations with the 100-m telescope of the MPIfR (Max-Planck-Institut f\"ur Radioastronomie) at Effelsberg.

\bibliographystyle{aa}  %
\bibliography{references} 
\begin{appendix}
\appendix
\section{Timing results from a 14-order DM polynomial\label{sec:appendix}}

In the main text, we present the results of the timing analysis of PSR J1913 + 1102 by modeling the DM variations with a Gaussian process and using the measured PK parameters to determine the component masses of the system and perform gravity tests.
As described in Sec.\,\ref{sec:obs}, to verify our results, we also adopt an alternative approach in which the DM variations are modeled using a DM polynomial. 

\begin{figure}[htbp]
    \centering
    \includegraphics[width=0.49\textwidth]{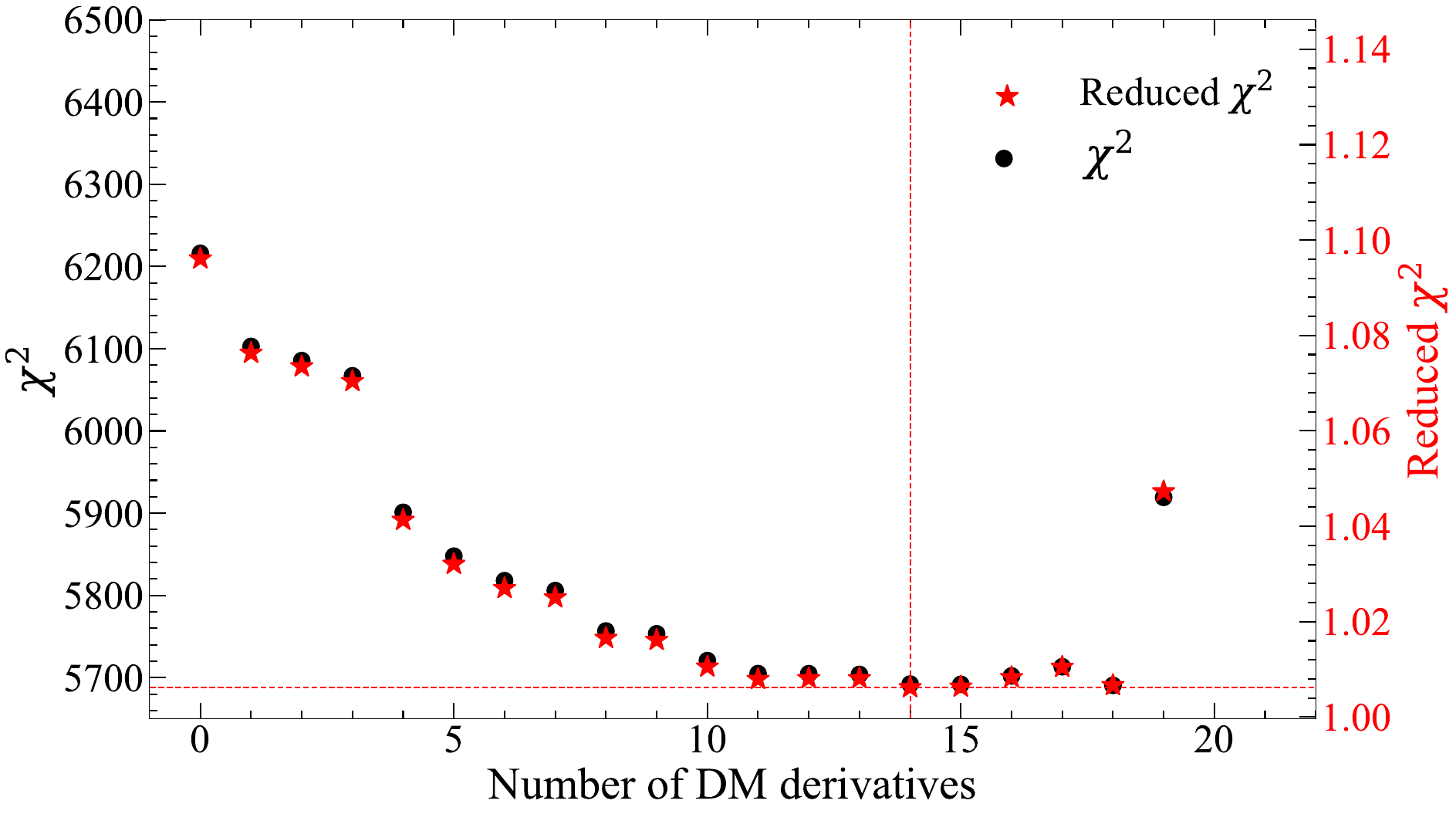}
    \caption{Variation of $\chi^{2}$ and reduced $\chi^{2}$ with the number of DM derivatives used in the fit. The 14-order derivative gives the minimum values, while higher orders lead to overfitting. \label{fig:DMderivatives}}
\end{figure}

In this method, to determine the optimal number of derivatives, we performed fits with successively higher derivative orders and recorded the corresponding values of $\chi^{2}$ and reduced $\chi^{2}$.
Fig.\,\ref{fig:DMderivatives} shows how $\chi^{2}$ and reduced $\chi^{2}$ vary with the number of DM derivatives. The results indicate that a 14-order DM derivative fit yields the lowest $\chi^{2}$ and reduced $\chi^{2}$, and higher orders degrade the fit quality, implying overfitting. The resulting fitted DM variation is shown in the bottom panel of Fig.\,\ref{fig:dmvariation}. In this figure, prior to MJD 57000, the 14-order DM polynomial model displays a steeper slope than the Gaussian process model, leading to slightly different timing results compared to the Gaussian process.

Using the 14-order DM polynomial method, the timing results are consistent at the 1-$\sigma$ level with those obtained from the Gaussian process. As an example, using the DDFWHE binary model, we get $\dot{\omega}=5.6511(3)\,{\rm deg\,yr^{-1}}$ and $\gamma=0.478(4)\,{\rm ms}$, from which we derive $m_{\rm tot}=2.88943(20)\,{\rm M_{\odot}}$, $m_{\rm p}=1.603(8)\,{\rm M_{\odot}}$, $m_{\rm c}=1.286(8)\,{\rm M_{\odot}}$ and $q=m_{\rm c}/m_{\rm p}=0.803(9)$. 
These are all 1-$\sigma$ consistent with the results of the DM fit in a Gaussian process. 
We employed an MCMC approach to compute the $\chi^{2}$ map over different values of the total mass $m_{\rm tot}$ and $\cos i$ in this method, and 
we get $\cos{i}=0.584(7)$, $m_{\rm c}=1.278(8)\,{\rm M_{\odot}}$, $m_{\rm p}=1.603(8)\,{\rm M_{\odot}}$ from these probability distributions, and they are consistent with results of DDFWHE model.

The most noticeable difference of the 14-order DM polynomial timing model is the proper motion: its magnitude is now smaller, $\mu=7.0(3)\,{\rm mas\,yr^{-1}}$. This is primarily due to the reduced declination component, $\mu_{\delta}=-6.23(35)\,{\rm mas\,yr^{-1}}$. Both values, listed in Table \,\ref{tab:extralpbdot}, are smaller by approximately 1 $\sigma$ compared to those derived from the Gaussian process. This is understandable as the proper motion is one of the effects most sensitive to long-term drift in the timing caused by the variations of the DM.

In the same table we list the kinematic contributions from the Shklovskii effect $\dot{P}_{\rm b}^{\rm Shk}$, the total non-intrinsic contributions $\dot{P}_{\rm b}^{\rm ext}$, the intrinsic contributions $\dot{P}_{\rm b}^{\rm intr}$, and the GR prediction for the orbital decay $\dot{P}_{\rm b}^{\rm GR}$, the latter being calculated with the mass values mentioned in the previous paragraph.
The contribution from differential Galactic acceleration, $\dot{P}_{\rm b}^{\rm Gal}$, is adopted to be the same as that listed in Tab.\,\ref{tab:extralpbdot_dmgp}.
The smaller proper motion leads to a weaker Shklovskii effect, while the observed orbital decay is slightly larger than that inferred from the Gaussian process.
Consequently, in the three models, we obtain a less negative value of $\dot{P}_{\rm b}^{\rm intr} = -453(6)\,{\rm fs\,s^{-1}}$, which is 1.5 $\sigma$ larger than $\dot{P}_{\rm b}^{\rm GR} = -461.8(5) \,{\rm fs\,s^{-1}}$. An alternative way to say this is that in this approach, the PK parameters $\dot{\omega}$, $\dot{P}_{\rm b}$, and $\gamma$ do not overlap within their 1-$\sigma$ uncertainties, owing to the intrinsic value of $\dot{P}_{\rm b}^{\rm intr}$ being less negative than that obtained with the Gaussian process. Nevertheless, within their 2-$\sigma$ uncertainties, all four PK parameters still agree in a region of the mass-mass diagram.

\renewcommand{\arraystretch}{1.3}
\begin{table}
\centering
\caption{We present the measured proper motion in right ascension $\mu_{\alpha}$ and in declination $\mu_{\delta}$, and the total proper motion $\mu$ from the 14-order DM polynomial. We list the shklovskii effect ($\dot{P}_{\rm b}^{\rm Shk}$),  the total non-intrinsic contributions ($\dot{P}_{\rm b}^{\rm ext}$) and the intrinsic contributions ($\dot{P}_{\rm b}^{\rm intr}$), all of which are affected by the measured proper motion.  We also list the GR prediction for the orbital decay ($\dot{P}_{\rm b}^{\rm GR}$) and the measured orbital decay ($\dot{P}_{\rm b}^{\rm obs}$). \label{tab:extralpbdot}}
\begin{tabular}{crrr}
\hline\hline
  \multicolumn{4}{c}{Measured parameters} \\
\hline
  $\mu_{\alpha}$ ($\rm mas\,yr^{-1}$)  & \multicolumn{3}{c}{$-3.27(16)$} \\
 $\mu_{\delta}$ ($\rm mas\,yr^{-1}$)  & \multicolumn{3}{c}{$-6.23(35)$} \\
 $\mu$ ($\rm mas\,yr^{-1}$)  & \multicolumn{3}{c}{$7.0(3)$} \\
$\dot P_{\rm b}^{\rm obs}$ ($\mathrm{fs\;s^{-1}}$) & \multicolumn{3}{c}{$-450.9(6.0)$} \\
\hline\hline
~ & NE2001 & YMW16 & NE2025\\
\hline
$\dot{P}_{\rm b}^{\rm Shk}\,({\rm fs\,s^{-1}})$ & $16.4^{+3.8}_{-3.5}$ & $15.3^{+3.5}_{-3.3}$ & $18.2^{+4.1}_{-3.9}$\\
$\dot{P}_{\rm b}^{\rm ext}\,({\rm fs\,s^{-1}})$ & $2.2^{+1.9}_{-1.8}$&  $2.2^{+1.8}_{-1.8}$ & $2.2^{+2.2}_{-1.9}$\\
$\dot{P}_{\rm b}^{\rm intr}\,({\rm fs\,s^{-1}})$ & $-453.1^{+6.2}_{-6.2}$ & $-453.1^{+6.2}_{-6.2}$ & $-453.3^{+6.3}_{-6.4}$\\
$\dot{P}_{\rm b}^{\rm GR}\,({\rm fs\,s^{-1}})$ & \multicolumn{3}{c}{$-461.8^{+0.5}_{-0.5}$}  \\
\hline
\end{tabular}
\end{table}

\end{appendix}
%
%

\end{document}